\documentclass[lettersize,journal]{IEEEtran}

\usepackage{circuitikz}
\usetikzlibrary{shapes,arrows}
\usetikzlibrary{decorations.pathreplacing}
\usetikzlibrary{math} 

\usepackage{cite}
\usepackage[T1]{fontenc}
\usepackage{graphicx}
\usepackage{amssymb}
\usepackage{amsmath} 
\usepackage{amsthm}
\usepackage{verbatim}
\usepackage{paralist}
\usepackage{setspace}
\usepackage{microtype}
\usepackage{hyperref}
\usepackage{url}
\usepackage{balance}
\usepackage{graphics}
\usepackage{dsfont}

\usepackage{multicol}

\usepackage{algorithm,algorithmic}
\usepackage{makecell}
\usepackage{overpic}
\usepackage{diagbox}
\usepackage{subcaption}

\newtheorem{theorem}{Theorem}
\newtheorem{corollary}{Corollary}

\newtheorem{remark}{Remark}

\newcommand{\blue}[1]{{\textcolor[rgb]{0,0,0.59}{#1}}}

\usepackage{comment}

\usepackage{amssymb}
\usepackage{amsfonts}
\usepackage{mathrsfs}
\usepackage{xspace}
\usepackage{bm}
\usepackage{upgreek}

\newcommand{\safemath}[2]{\newcommand{#1}{\ensuremath{#2}\xspace}}

\safemath{\bma}{\mathbf{a}}
\safemath{\bmb}{\mathbf{b}}
\safemath{\bmc}{\mathbf{c}}
\safemath{\bmd}{\mathbf{d}}
\safemath{\bme}{\mathbf{e}}
\safemath{\bmf}{\mathbf{f}}
\safemath{\bmg}{\mathbf{g}}
\safemath{\bmh}{\mathbf{h}}
\safemath{\bmi}{\mathbf{i}}
\safemath{\bmj}{\mathbf{j}}
\safemath{\bmk}{\mathbf{k}}
\safemath{\bml}{\mathbf{l}}
\safemath{\bmm}{\mathbf{m}}
\safemath{\bmn}{\mathbf{n}}
\safemath{\bmo}{\mathbf{o}}
\safemath{\bmp}{\mathbf{p}}
\safemath{\bmq}{\mathbf{q}}
\safemath{\bmr}{\mathbf{r}}
\safemath{\bms}{\mathbf{s}}
\safemath{\bmt}{\mathbf{t}}
\safemath{\bmu}{\mathbf{u}}
\safemath{\bmv}{\mathbf{v}}
\safemath{\bmw}{\mathbf{w}}
\safemath{\bmx}{\mathbf{x}}
\safemath{\bmy}{\mathbf{y}}
\safemath{\bmz}{\mathbf{z}}
\safemath{\bmzero}{\mathbf{0}}
\safemath{\bmone}{\mathbf{1}}

\bmdefine{\biad}{a}
\bmdefine{\bibd}{b}
\bmdefine{\bicd}{c}
\bmdefine{\bidd}{d}
\bmdefine{\bied}{e}
\bmdefine{\bifd}{f}
\bmdefine{\bigd}{g}
\bmdefine{\bihd}{h}
\bmdefine{\biid}{i}
\bmdefine{\bijd}{j}
\bmdefine{\bikd}{k}
\bmdefine{\bild}{l}
\bmdefine{\bimd}{m}
\bmdefine{\bind}{n}
\bmdefine{\biod}{o}
\bmdefine{\bipd}{p}
\bmdefine{\biqd}{q}
\bmdefine{\bird}{r}
\bmdefine{\bisd}{s}
\bmdefine{\bitd}{t}
\bmdefine{\biud}{u}
\bmdefine{\bivd}{v}
\bmdefine{\biwd}{w}
\bmdefine{\bixd}{x}
\bmdefine{\biyd}{y}
\bmdefine{\bizd}{z}

\bmdefine{\bixid}{\xi}
\bmdefine{\bilambdad}{\lambda}
\bmdefine{\bimud}{\mu}
\bmdefine{\bithetad}{\theta}
\bmdefine{\biphid}{\phi}
\bmdefine{\bideltad}{\delta}

\safemath{\bmia}{\biad}
\safemath{\bmib}{\bibd}
\safemath{\bmic}{\bicd}
\safemath{\bmid}{\bidd}
\safemath{\bmie}{\bied}
\safemath{\bmif}{\bifd}
\safemath{\bmig}{\bigd}
\safemath{\bmih}{\bihd}
\safemath{\bmii}{\biid}
\safemath{\bmij}{\bijd}
\safemath{\bmik}{\bikd}
\safemath{\bmil}{\bild}
\safemath{\bmim}{\bimd}
\safemath{\bmin}{\bind}
\safemath{\bmio}{\biod}
\safemath{\bmip}{\bipd}
\safemath{\bmiq}{\biqd}
\safemath{\bmir}{\bird}
\safemath{\bmis}{\bisd}
\safemath{\bmit}{\bitd}
\safemath{\bmiu}{\biud}
\safemath{\bmiv}{\bivd}
\safemath{\bmiw}{\biwd}
\safemath{\bmix}{\bixd}
\safemath{\bmiy}{\biyd}
\safemath{\bmiz}{\bizd}

\safemath{\bmxi}{\bixid}
\safemath{\bmlambda}{\bilambdad}
\safemath{\bmmu}{\bimud}
\safemath{\bmtheta}{\bithetad}
\safemath{\bmphi}{\biphid}
\safemath{\bmdelta}{\bideltad}

\safemath{\bA}{\mathbf{A}}
\safemath{\bB}{\mathbf{B}}
\safemath{\bC}{\mathbf{C}}
\safemath{\bD}{\mathbf{D}}
\safemath{\bE}{\mathbf{E}}
\safemath{\bF}{\mathbf{F}}
\safemath{\bG}{\mathbf{G}}
\safemath{\bH}{\mathbf{H}}
\safemath{\bI}{\mathbf{I}}
\safemath{\bJ}{\mathbf{J}}
\safemath{\bK}{\mathbf{K}}
\safemath{\bL}{\mathbf{L}}
\safemath{\bM}{\mathbf{M}}
\safemath{\bN}{\mathbf{N}}
\safemath{\bO}{\mathbf{O}}
\safemath{\bP}{\mathbf{P}}
\safemath{\bQ}{\mathbf{Q}}
\safemath{\bR}{\mathbf{R}}
\safemath{\bS}{\mathbf{S}}
\safemath{\bT}{\mathbf{T}}
\safemath{\bU}{\mathbf{U}}
\safemath{\bV}{\mathbf{V}}
\safemath{\bW}{\mathbf{W}}
\safemath{\bX}{\mathbf{X}}
\safemath{\bY}{\mathbf{Y}}
\safemath{\bZ}{\mathbf{Z}}

\safemath{\bZero}{\mathbf{0}}
\safemath{\bOne}{\mathbf{1}}
\safemath{\bDelta}{\mathbf{\Delta}}
\safemath{\bLambda}{\mathbf{\UpLambda}}
\safemath{\bPhi}{\mathbf{\Upphi}}
\safemath{\bSigma}{\mathbf{\Upsigma}}
\safemath{\bOmega}{\mathbf{\Upomega}}
\safemath{\bTheta}{\mathbf{\Uptheta}}

\bmdefine{\biAd}{A}
\bmdefine{\biBd}{B}
\bmdefine{\biCd}{C}
\bmdefine{\biDd}{D}
\bmdefine{\biEd}{E}
\bmdefine{\biFd}{F}
\bmdefine{\biGd}{G}
\bmdefine{\biHd}{H}
\bmdefine{\biId}{I}
\bmdefine{\biJd}{J}
\bmdefine{\biKd}{K}
\bmdefine{\biLd}{L}
\bmdefine{\biMd}{M}
\bmdefine{\biOd}{N}
\bmdefine{\biPd}{O}
\bmdefine{\biQd}{P}
\bmdefine{\biRd}{R}
\bmdefine{\biSd}{S}
\bmdefine{\biTd}{T}
\bmdefine{\biUd}{U}
\bmdefine{\biVd}{V}
\bmdefine{\biWd}{W}
\bmdefine{\biXd}{X}
\bmdefine{\biYd}{Y}
\bmdefine{\biZd}{Z}

\bmdefine{\biDelta}{\Delta}
\bmdefine{\biLambda}{\Lambda}
\bmdefine{\biPhi}{\Phi}
\bmdefine{\biSigma}{\Sigma}
\bmdefine{\biOmega}{\Omega}
\bmdefine{\biTheta}{\Theta}

\safemath{\bimA}{\biAd}
\safemath{\bimB}{\biBd}
\safemath{\bimC}{\biCd}
\safemath{\bimD}{\biDd}
\safemath{\bimE}{\biEd}
\safemath{\bimF}{\biFd}
\safemath{\bimG}{\biGd}
\safemath{\bimH}{\biHd}
\safemath{\bimI}{\biId}
\safemath{\bimJ}{\biJd}
\safemath{\bimK}{\biKd}
\safemath{\bimL}{\biLd}
\safemath{\bimM}{\biMd}
\safemath{\bimN}{\biNd}
\safemath{\bimO}{\biOd}
\safemath{\bimP}{\biPd}
\safemath{\bimQ}{\biQd}
\safemath{\bimR}{\biRd}
\safemath{\bimS}{\biSd}
\safemath{\bimT}{\biTd}
\safemath{\bimU}{\biUd}
\safemath{\bimV}{\biVd}
\safemath{\bimW}{\biWd}
\safemath{\bimX}{\biXd}
\safemath{\bimY}{\biYd}
\safemath{\bimZ}{\biZd}

\safemath{\bimDelta}{\biDelta}
\safemath{\bimLambda}{\biLambda}
\safemath{\bimPhi}{\biPhi}
\safemath{\bimSigma}{\biSigma}
\safemath{\bimOmega}{\biOmega}
\safemath{\bimTheta}{\biTheta}

\safemath{\setA}{\mathcal{A}}
\safemath{\setB}{\mathcal{B}}
\safemath{\setC}{\mathcal{C}}
\safemath{\setD}{\mathcal{D}}
\safemath{\setE}{\mathcal{E}}
\safemath{\setF}{\mathcal{F}}
\safemath{\setG}{\mathcal{G}}
\safemath{\setH}{\mathcal{H}}
\safemath{\setI}{\mathcal{I}}
\safemath{\setJ}{\mathcal{J}}
\safemath{\setK}{\mathcal{K}}
\safemath{\setL}{\mathcal{L}}
\safemath{\setM}{\mathcal{M}}
\safemath{\setN}{\mathcal{N}}
\safemath{\setO}{\mathcal{O}}
\safemath{\setP}{\mathcal{P}}
\safemath{\setQ}{\mathcal{Q}}
\safemath{\setR}{\mathcal{R}}
\safemath{\setS}{\mathcal{S}}
\safemath{\setT}{\mathcal{T}}
\safemath{\setU}{\mathcal{U}}
\safemath{\setV}{\mathcal{V}}
\safemath{\setW}{\mathcal{W}}
\safemath{\setX}{\mathcal{X}}
\safemath{\setY}{\mathcal{Y}}
\safemath{\setZ}{\mathcal{Z}}
\safemath{\emptySet}{\varnothing}

\safemath{\colA}{\mathscr{A}}
\safemath{\colB}{\mathscr{B}}
\safemath{\colC}{\mathscr{C}}
\safemath{\colD}{\mathscr{D}}
\safemath{\colE}{\mathscr{E}}
\safemath{\colF}{\mathscr{F}}
\safemath{\colG}{\mathscr{G}}
\safemath{\colH}{\mathscr{H}}
\safemath{\colI}{\mathscr{I}}
\safemath{\colJ}{\mathscr{J}}
\safemath{\colK}{\mathscr{K}}
\safemath{\colL}{\mathscr{L}}
\safemath{\colM}{\mathscr{M}}
\safemath{\colN}{\mathscr{N}}
\safemath{\colO}{\mathscr{O}}
\safemath{\colP}{\mathscr{P}}
\safemath{\colQ}{\mathscr{Q}}
\safemath{\colR}{\mathscr{R}}
\safemath{\colS}{\mathscr{S}}
\safemath{\colT}{\mathscr{T}}
\safemath{\colU}{\mathscr{U}}
\safemath{\colV}{\mathscr{V}}
\safemath{\colW}{\mathscr{W}}
\safemath{\colX}{\mathscr{X}}
\safemath{\colY}{\mathscr{Y}}
\safemath{\colZ}{\mathscr{Z}}

\safemath{\opA}{\mathbb{A}}
\safemath{\opB}{\mathbb{B}}
\safemath{\opC}{\mathbb{C}}
\safemath{\opD}{\mathbb{D}}
\safemath{\opE}{\mathbb{E}}
\safemath{\opF}{\mathbb{F}}
\safemath{\opG}{\mathbb{G}}
\safemath{\opH}{\mathbb{H}}
\safemath{\opI}{\mathbb{I}}
\safemath{\opJ}{\mathbb{J}}
\safemath{\opK}{\mathbb{K}}
\safemath{\opL}{\mathbb{L}}
\safemath{\opM}{\mathbb{M}}
\safemath{\opN}{\mathbb{N}}
\safemath{\opO}{\mathbb{O}}
\safemath{\opP}{\mathbb{P}}
\safemath{\opQ}{\mathbb{Q}}
\safemath{\opR}{\mathbb{R}}
\safemath{\opS}{\mathbb{S}}
\safemath{\opT}{\mathbb{T}}
\safemath{\opU}{\mathbb{U}}
\safemath{\opV}{\mathbb{V}}
\safemath{\opW}{\mathbb{W}}
\safemath{\opX}{\mathbb{X}}
\safemath{\opY}{\mathbb{Y}}
\safemath{\opZ}{\mathbb{Z}}
\safemath{\opZero}{\mathbb{O}}
\safemath{\identityop}{\opI}

\safemath{\veca}{\bma}
\safemath{\vecb}{\bmb}
\safemath{\vecc}{\bmc}
\safemath{\vecd}{\bmd}
\safemath{\vece}{\bme}
\safemath{\vecf}{\bmf}
\safemath{\vecg}{\bmg}
\safemath{\vech}{\bmh}
\safemath{\veci}{\bmi}
\safemath{\vecj}{\bmj}
\safemath{\veck}{\bmk}
\safemath{\vecl}{\bml}
\safemath{\vecm}{\bmm}
\safemath{\vecn}{\bmn}
\safemath{\veco}{\bmo}
\safemath{\vecp}{\bmp}
\safemath{\vecq}{\bmq}
\safemath{\vecr}{\bmr}
\safemath{\vecs}{\bms}
\safemath{\vect}{\bmt}
\safemath{\vecu}{\bmu}
\safemath{\vecv}{\bmv}
\safemath{\vecw}{\bmw}
\safemath{\vecx}{\bmx}
\safemath{\vecy}{\bmy}
\safemath{\vecz}{\bmz}

\safemath{\veczero}{\bmzero}
\safemath{\vecone}{\bmone}
\safemath{\vecxi}{\bmxi}
\safemath{\veclambda}{\bmlambda}
\safemath{\vecmu}{\bmmu}
\safemath{\vectheta}{\bmtheta}
\safemath{\vecphi}{\bmphi}
\safemath{\vecdelta}{\bmdelta}

\safemath{\matA}{\bA}
\safemath{\matB}{\bB}
\safemath{\matC}{\bC}
\safemath{\matD}{\bD}
\safemath{\matE}{\bE}
\safemath{\matF}{\bF}
\safemath{\matG}{\bG}
\safemath{\matH}{\bH}
\safemath{\matI}{\bI}
\safemath{\matJ}{\bJ}
\safemath{\matK}{\bK}
\safemath{\matL}{\bL}
\safemath{\matM}{\bM}
\safemath{\matN}{\bN}
\safemath{\matO}{\bO}
\safemath{\matP}{\bP}
\safemath{\matQ}{\bQ}
\safemath{\matR}{\bR}
\safemath{\matS}{\bS}
\safemath{\matT}{\bT}
\safemath{\matU}{\bU}
\safemath{\matV}{\bV}
\safemath{\matW}{\bW}
\safemath{\matX}{\bX}
\safemath{\matY}{\bY}
\safemath{\matZ}{\bZ}
\safemath{\matzero}{\bmzero}

\safemath{\matDelta}{\bDelta}
\safemath{\matLambda}{\bLambda}
\safemath{\matPhi}{\bPhi}
\safemath{\matSigma}{\bSigma}
\safemath{\matOmega}{\bOmega}
\safemath{\matTheta}{\bTheta}

\safemath{\matidentity}{\matI}
\safemath{\matone}{\matO}

\safemath{\rnda}{A}
\safemath{\rndb}{B}
\safemath{\rndc}{C}
\safemath{\rndd}{D}
\safemath{\rnde}{E}
\safemath{\rndf}{F}
\safemath{\rndg}{G}
\safemath{\rndh}{H}
\safemath{\rndi}{I}
\safemath{\rndj}{J}
\safemath{\rndk}{K}
\safemath{\rndl}{L}
\safemath{\rndm}{M}
\safemath{\rndn}{N}
\safemath{\rndo}{O}
\safemath{\rndp}{P}
\safemath{\rndq}{Q}
\safemath{\rndr}{R}
\safemath{\rnds}{S}
\safemath{\rndt}{T}
\safemath{\rndu}{U}
\safemath{\rndv}{V}
\safemath{\rndw}{W}
\safemath{\rndx}{X}
\safemath{\rndy}{Y}
\safemath{\rndz}{Z}

\safemath{\rveca}{\bimA}
\safemath{\rvecb}{\bimB}
\safemath{\rvecc}{\bimC}
\safemath{\rvecd}{\bimD}
\safemath{\rvece}{\bimE}
\safemath{\rvecf}{\bimF}
\safemath{\rvecg}{\bimG}
\safemath{\rvech}{\bimH}
\safemath{\rveci}{\bimI}
\safemath{\rvecj}{\bimJ}
\safemath{\rveck}{\bimK}
\safemath{\rvecl}{\bimL}
\safemath{\rvecm}{\bimM}
\safemath{\rvecn}{\bimN}
\safemath{\rveco}{\bomO}
\safemath{\rvecp}{\bimP}
\safemath{\rvecq}{\bimQ}
\safemath{\rvecr}{\bimR}
\safemath{\rvecs}{\bimS}
\safemath{\rvect}{\bimT}
\safemath{\rvecu}{\bimU}
\safemath{\rvecv}{\bimV}
\safemath{\rvecw}{\bimW}
\safemath{\rvecx}{\bimX}
\safemath{\rvecy}{\bimY}
\safemath{\rvecz}{\bimZ}

\safemath{\rvecxi}{\bmxi}
\safemath{\rveclambda}{\bmlambda}
\safemath{\rvecmu}{\bmmu}
\safemath{\rvectheta}{\bmtheta}
\safemath{\rvecphi}{\bmphi}

\safemath{\rmatA}{\bimA}
\safemath{\rmatB}{\bimB}
\safemath{\rmatC}{\bimC}
\safemath{\rmatD}{\bimD}
\safemath{\rmatE}{\bimE}
\safemath{\rmatF}{\bimF}
\safemath{\rmatG}{\bimG}
\safemath{\rmatH}{\bimH}
\safemath{\rmatI}{\bimI}
\safemath{\rmatJ}{\bimJ}
\safemath{\rmatK}{\bimK}
\safemath{\rmatL}{\bimL}
\safemath{\rmatM}{\bimM}
\safemath{\rmatN}{\bimN}
\safemath{\rmatO}{\bimO}
\safemath{\rmatP}{\bimP}
\safemath{\rmatQ}{\bimQ}
\safemath{\rmatR}{\bimR}
\safemath{\rmatS}{\bimS}
\safemath{\rmatT}{\bimT}
\safemath{\rmatU}{\bimU}
\safemath{\rmatV}{\bimV}
\safemath{\rmatW}{\bimW}
\safemath{\rmatX}{\bimX}
\safemath{\rmatY}{\bimY}
\safemath{\rmatZ}{\bimZ}

\safemath{\rmatDelta}{\bimDelta}
\safemath{\rmatLambda}{\bimLambda}
\safemath{\rmatPhi}{\bimPhi}
\safemath{\rmatSigma}{\bimSigma}
\safemath{\rmatOmega}{\bimOmega}
\safemath{\rmatTheta}{\bimTheta}

\usepackage{amssymb}
\usepackage{amsfonts}
\usepackage{mathrsfs}
\usepackage{xspace}
\usepackage{bm}
\usepackage{fancyref}
\usepackage{textcomp}

\usepackage{multirow}
\usepackage{stmaryrd}

\newenvironment{textbmatrix}{	\setlength{\arraycolsep}{2.5pt}%
								\big[\begin{matrix}}{\end{matrix}\big]%
								\raisebox{0.08ex}{\vphantom{M}}}

\def\be{\begin{equation}}
\def\ee{\end{equation}}
\def\een{\nonumber \end{equation}}
\def\mat{\begin{bmatrix}}
\def\emat{\end{bmatrix}}
\def\btm{\begin{textbmatrix}}
\def\etm{\end{textbmatrix}}

\def\ba#1\ea{\begin{align}#1\end{align}}
\def\bas#1\eas{\begin{align*}#1\end{align*}}
\def\bs#1\es{\begin{split}#1\end{split}} 
\def\bg#1\eg{\begin{gather}#1\end{gather}}
\def\bml#1\eml{\begin{multline}#1\end{multline}}
\def\bi#1\ei{\begin{itemize}#1\end{itemize}}

\safemath{\dirac}{\delta}					
\safemath{\krond}{\dirac}					

\safemath{\upto}{\uparrow}
\safemath{\downto}{\downarrow}
\safemath{\iu}{j}							
\safemath{\ev}{\lambda}						
\safemath{\hilseqspace}{l^{2}}				
\newcommand{\banachfunspace}[1]{\setL^{#1}}	
\safemath{\hilfunspace}{\banachfunspace{2}}	

\safemath{\SNR}{\textsf{SNR}} 				
\safemath{\PAR}{\textsf{PAR}} 				
\safemath{\No}{N_0}							
\safemath{\Es}{E_s}							
\safemath{\Eb}{E_b}							
\safemath{\EbNo}{\frac{\Eb}{\No}}
\safemath{\EsNo}{\frac{\Es}{\No}}

\DeclareMathOperator{\CHop}{\ensuremath{\opH}} 
\safemath{\tvir}{\rndh_{\CHop}}				
\safemath{\tvtf}{\rndl_{\CHop}}				
\safemath{\spf}{\rnds_{\CHop}}				
\safemath{\bff}{H_{\CHop}}					

\safemath{\ircf}{r_{h}}						
\safemath{\tftvcf}{r_{s}}					
\safemath{\tfcf}{r_{l}}						
\safemath{\bfcf}{r_{H}}						

\safemath{\tcorr}{c_h}						
\safemath{\scf}{c_{s}}						
\safemath{\tfcorr}{c_{l}}					
\safemath{\fcorr}{c_{H}}						

\safemath{\mi}{I}							
\safemath{\capacity}{C}						

\safemath{\normal}{\mathcal{N}}			
\safemath{\jpg}{\mathcal{CN}}			
\safemath{\mchain}{\leftrightarrow}		

\safemath{\dB}{\,\mathrm{dB}}
\safemath{\dBm}{\,\mathrm{dBm}}
\safemath{\Hz}{\,\mathrm{Hz}}
\safemath{\kHz}{\,\mathrm{kHz}}
\safemath{\MHz}{\,\mathrm{MHz}}
\safemath{\GHz}{\,\mathrm{GHz}}
\safemath{\s}{\,\mathrm{s}}
\safemath{\ms}{\,\mathrm{ms}}
\safemath{\mus}{\,\mathrm{\text{\textmu}s}}
\safemath{\ns}{\,\mathrm{ns}}
\safemath{\ps}{\,\mathrm{ps}}
\safemath{\meter}{\,\mathrm{m}}
\safemath{\mm}{\,\mathrm{mm}}
\safemath{\cm}{\,\mathrm{cm}}
\safemath{\m}{\,\mathrm{m}}
\safemath{\W}{\,\mathrm{W}}
\safemath{\mW}{\, \mathrm{mW}}
\safemath{\J}{\,\mathrm{J}}
\safemath{\K}{\,\mathrm{K}}
\safemath{\bit}{\,\mathrm{bit}}
\safemath{\nat}{\,\mathrm{nat}}

\safemath{\define}{\triangleq}			

\safemath{\equivalent}{\sim}
\safemath{\distas}{\sim}					
\safemath{\sdiff}{\Delta}				

\safemath{\reals}{\mathbb{R}}
\safemath{\positivereals}{\reals_{+}}
\safemath{\integers}{\mathbb{Z}}
\safemath{\posint}{\integers_{+}}
\safemath{\naturals}{\mathbb{N}}
\safemath{\posnaturals}{\naturals_{+}}
\safemath{\complexset}{\mathbb{C}}
\safemath{\rationals}{\mathbb{Q}}

\newcommand*{\fancyrefapplabelprefix}{app}		
\newcommand*{\fancyrefthmlabelprefix}{thm}		
\newcommand*{\fancyreflemlabelprefix}{lem}		
\newcommand*{\fancyrefcorlabelprefix}{cor}		
\newcommand*{\fancyrefdeflabelprefix}{def}		
\newcommand*{\fancyrefproplabelprefix}{prop}	
\newcommand*{\fancyrefobslabelprefix}{obs}		
\newcommand*{\fancyrefalglabelprefix}{alg}		
\newcommand*{\fancyrefasmlabelprefix}{asm}	    

\frefformat{vario}{\fancyrefseclabelprefix}{Section~#1}
\frefformat{vario}{\fancyrefthmlabelprefix}{Theorem~#1}
\frefformat{vario}{\fancyreflemlabelprefix}{Lemma~#1}
\frefformat{vario}{\fancyrefcorlabelprefix}{Corollary~#1}
\frefformat{vario}{\fancyrefdeflabelprefix}{Definition~#1}
\frefformat{vario}{\fancyrefobslabelprefix}{Observation~#1}
\frefformat{vario}{\fancyrefasmlabelprefix}{Assumption~#1}
\frefformat{vario}{\fancyreffiglabelprefix}{Fig.~#1}
\frefformat{vario}{\fancyrefapplabelprefix}{Appendix~#1} 
\frefformat{vario}{\fancyrefproplabelprefix}{Proposition~#1}
\frefformat{vario}{\fancyrefalglabelprefix}{Algorithm~#1}
\frefformat{vario}{\fancyrefeqlabelprefix}{(#1)}

\safemath{\dictab}{[\,\dicta\,\,\dictb\,]}

\safemath{\ysig}{\bmy}
\safemath{\ysighat}{\hat{\ysig}}
\safemath{\ysigdim}{M}
\safemath{\xsig}{\bmx}
\safemath{\xsigdim}{N}
\safemath{\nx}{n_x}
\safemath{\zsig}{\bmz}
\safemath{\zsigdim}{\ysigdim}
\safemath{\rsig}{\bmr}
\safemath{\Adict}{\bA}
\safemath{\Adicttilde}{\widetilde{\Adict}}
\safemath{\Adictdim}{\outputdim\times\xsigdim}
\safemath{\avec}{\bma}
\safemath{\avectilde}{\tilde{\avec}}
\safemath{\Bdict}{\bB}
\safemath{\Bdicttilde}{\widetilde{\Bdict}}
\safemath{\Cdict}{\bC}
\safemath{\cvec}{\bmc}
\safemath{\Ddict}{\bD}
\safemath{\Ddictdim}{\ysigdim\times\xsigdim}
\safemath{\dvec}{\bmd}
\safemath{\Ddicttilde}{\widetilde{\bD}}
\safemath{\Bonb}{\bB}
\safemath{\bvec}{\bmb}
\safemath{\Bonbdim}{\ysigdim\times\ysigdim}
\safemath{\noise}{\bmn}
\safemath{\noisedim}{\ysigim}
\safemath{\err}{\bme}
\safemath{\errdim}{\ysigdim}
\safemath{\errset}{\setE}
\safemath{\nerr}{n_e}
\safemath{\delop}{\bP_\errset}
\safemath{\delopc}{\bP_{{\errset}^c}}

\safemath{\cplxi}{\imath}
\safemath{\cplxj}{\jmath}

\safemath{\dict}{\matD}
\safemath{\inputdim}{N}		
\safemath{\outputdim}{M}		
\safemath{\sparsity}{S}	
\safemath{\inputdimA}{{N_a}}	
\safemath{\inputdimB}{{N_b}}	
\safemath{\elemA}{{n_a}}	
\safemath{\elemB}{{n_b}}	
\safemath{\resA}{\matR_a}	
\safemath{\resB}{\matR_b}	
\safemath{\subD}{\matS} 
\safemath{\subA}{\matS_a} 
\safemath{\subB}{\matS_b} 
\safemath{\dicta}{\matA} 	
\safemath{\dictb}{\matB} 	
\safemath{\hollowS}{H}
\safemath{\hollowA}{H_a}
\safemath{\hollowB}{H_b}
\safemath{\cross}{Z}
\safemath{\coh}{\mu_d}			
\safemath{\coha}{\mu_a}			
\safemath{\cohb}{\mu_b}			
\safemath{\mubs}{\nu}	
\safemath{\cohm}{\mu_m} 
\safemath{\dictset}{\setD}	
\safemath{\dictsetp}{\dictset(\coh,\coha,\cohb)}	
\safemath{\dictsetgen}{\dictset_\text{gen}}
\safemath{\dictsetgenp}{\dictsetgen(\coh)}
\safemath{\dictsetonb}{\dictset_\text{onb}}
\safemath{\dictsetonbp}{\dictsetonb(\coh)}

\safemath{\leftside}{U}
\safemath{\rightsideA}{R_a}
\safemath{\rightsideB}{R_b}

\safemath{\indexS}{\setI_S} 

\safemath{\na}{n_a}			
\safemath{\nb}{n_b}			
\safemath{\coeffa}{p_i}	
\safemath{\coeffb}{q_j}	
\safemath{\seta}{\setP}		
\safemath{\setb}{\setQ}     
\safemath{\setw}{\setW}	
\safemath{\setz}{\setZ}	
\safemath{\cola}{\veca}		
\safemath{\colb}{\vecb}		
\safemath{\cold}{\vecd}		
\safemath{\inputvec}{\vecx} 	
\safemath{\error}{\vece}	
\safemath{\noiseout}{\vecz} 	
\safemath{\inputvecel}{x}
\safemath{\inputveca}{\vecx_a}
\safemath{\inputvecb}{\vecx_b}
\safemath{\outputvec}{\vecy}	
\safemath{\lambdamin}{\lambda_{\mathrm{min}}}

\safemath{\elltwo}{\ell_2}
\safemath{\ellone}{\ell_1}
\safemath{\ellzero}{\ell_0}
\safemath{\ellinf}{\ell_\infty}
\safemath{\ellinftilde}{\ell_{\widetilde\infty}}
\safemath{\licard}{Z(\coh,\coha,\cohb)}
\safemath{\xsol}{\hat{x}}
\safemath{\xbord}{x_b}		
\safemath{\xstat}{x_s}		
\safemath{\xstatLone}{\tilde{x}_s}
\safemath{\order}{\mathcal{O}} 
\safemath{\scales}{\Theta} 
\safemath{\ones}{\mathbf{1}} 
\safemath{\zeroes}{\mathbf{0}} 
\safemath{\thlone}{\kappa(\coh,\cohb)} 
\safemath{\constoneA}{\delta} 
\safemath{\constoneB}{\epsilon} 
\safemath{\nlarge}{L}				   
\safemath{\sumlarge}{S_\nlarge}
\safemath{\maxlarger}{P_\nlarge}	   
\safemath{\Pzero}{\textrm{P0}}	
\safemath{\Pone}{\textrm{P1}}
\safemath{\vecfir}{\vecw}			 
\safemath{\vecsec}{\vecz}
\safemath{\elvecfir}{w}              
\safemath{\elvecsec}{z}				 
\safemath{\nlargefir}{n}
\safemath{\normout}{\gamma}
\safemath{\auxfun}{h}
\safemath{\supp}{\textrm{supp}}

\safemath{\indexa}{\ell}
\safemath{\indexb}{r}
\safemath{\indexc}{i}
\safemath{\indexd}{j}

\safemath{\project}{P}

\def\diag{\mathrm{diag}}

\def\Htran{\mbox{\tiny $\mathrm{H}$}}

\def\SNR{\mbox{\footnotesize $\mathrm{SNR}$}}

\def\mod{\mathrm{mod}}

\ifCLASSINFOpdf
\else
\fi
\begin{document}
%
\title{Nonlinear Distortion Radiated from Large Arrays and Active Reconfigurable Intelligent Surfaces}
%
%
%

\author{Nikolaos~Kolomvakis, Alva Kosasih
        and~Emil~Bj\"{o}rnson,~\IEEEmembership{Fellow,~IEEE}
\thanks{N. Kolomvakis and E. Bj\"{o}rnson are with the Division of Communication Systems, KTH Royal Institute of Technology, 164 40
Stockholm, Sweden (e-mail:\{nikkol, emilbjo\}@kth.se).}
\thanks{A. Kosasih was with KTH Royal Institute of Technology, Stockholm, Sweden. He is now with Nokia, Espoo, Finland (e-mail: alva.kosasih@nokia.com).}
\thanks{This work was supported by the Knut and Alice Wallenberg
Foundation through the Wallenberg Academy Fellow program, and also by the SweWIN competence center funded by VINNOVA.}
\thanks{This article was presented in part at the European Conference on Antennas and Propagation (EuCAP), Glasgow, Scotland, 2024\cite{eucap2024}.}
}

\maketitle

\begin{abstract}
Extremely large aperture arrays (ELAAs) and reconfigurable intelligent surfaces (RISs) are candidate enablers to realize connectivity goals for the sixth-generation (6G) wireless networks. For instance, ELAAs can provide orders-of-magnitude higher area throughput compared to what massive multiple-input multiple-output (MIMO) can deliver through spatial multiplexing, while RISs can improve the propagation conditions over wireless channels but a passively reflecting RIS must be large to be effective. Active RIS with amplifiers can deal with this issue. In this paper, we analyze the distortion generated by nonlinear amplifiers in both ELAAs and active RIS. We derive analytical expressions for the angular directions and depth of nonlinear distortion in both near-field and far-field channels. These insights are then used in a distortion-aware scheduling scheme that predicts the beamforming directions of the distortion and strategically allocates users in frequency to minimize its impact. Numerical results validate our theoretical analysis and compare distortion-aware and distortion-unaware scheduling methods, highlighting the benefits of accounting for nonlinearities. We conclude that nonlinearities can both create in-band and out-of-band distortion that is beamformed in entirely new directions and distances from the transmitter.
\end{abstract}


%
\IEEEpeerreviewmaketitle

\section{Introduction}
%
%
%
%

\IEEEPARstart{E}{xtremely} large aperture arrays (ELAAs) and reconfigurable intelligent surfaces (RISs) are promising technologies for the next generation of wireless systems. These are electrically large arrays respectively used for transmission/reception and reflection.
The aim of deploying ELAAs is to provide orders-of-magnitude higher area throughput in wireless networks compared to what massive multiple-input multiple-output (MIMO) can practically deliver. The keys to reaching this goal are the even larger number of antennas and huge apertures that place the users in the radiative near-field, which reduces the average propagation loss and increases the spatial resolution by controlling beams in the depth domain \cite{Bjornson2019a}. 

On the other hand, RISs can turn wireless propagation channels into programmable smart environments by software-controlled reflection of impinging signals \cite{Huang2018a,RIS_Renzo_JSAC}. The aim can be to boost the received signal power at the receiver and improve other desired channel properties such as the MIMO channel rank. The basic idea is to densify the next generation of wireless networks with RISs instead of base stations (BSs) to create a low-cost, low-power, and high-throughput communication infrastructure. This new architecture is characterized by being programmable and reconfigurable, not only at the infrastructure nodes and user devices, but also in the surrounding wireless environment, properly shaped by programmable RISs. Each RIS acts as an advanced kind of transparent full-duplex relay/repeater \cite{ris_myths,BJornson2020, DiRenzo2020}.
Each unit cell of a RIS can be made of metamaterial, acts as an isotropic scatterer when it is sub-wavelength-sized, and the impedance can be tuned by diodes to create an inhomogeneous phase-shift pattern over the surface that reflects an incident wave as a beam in a desirable direction \cite{Wu2021a}. 

A conventional \emph{passive RIS} reflects signals without amplification. This has the benefit of low power consumption but requires a large aperture size since the pathloss is the product of the pathlosses to and from the array, while the array gain grows quadratically with the aperture area \cite{Wu2021a}. Nevertheless, in practice, the capacity gains are typically only substantial in setups where the direct link between transmitter and receiver is completely blocked or very weak \cite{Bjornson2022a}. 
To expand the usefulness of the RIS technology to scenarios where the direct link is not weak, in \cite{active_ris}, a new RIS architecture called \emph{active RISs} was proposed to overcome the large path loss of reflected links. 
The key feature of active RISs is their ability to actively reflect signals with amplification. This can be realized by integrating a reflection-type amplifier into each reflecting cell. Different from conventional relays, an active RIS remains to be transparent and full duplex.

The use of low-cost, small-size, and power-efficient hardware is desirable to make ELAA and active RISs sustainable technological shifts, otherwise the cost will increase with the number of antennas. However, decreasing the hardware cost and size as well as increasing its power efficiency typically implies various radio-frequency (RF) imperfections. This issue has been raising increasing interest in the last decade with the advent of massive MIMO systems \cite{Nikos2016,Nikos2017,mollen_spatial,nikos_quant}, including, e.g., the impacts of quantization noise in converters, oscillator phase noise, transceiver I/Q imbalances, and amplifier nonlinearities.
For example, practical power amplifiers (PAs) introduce both in-band and out-of-band distortion due to their nonlinear transfer function. This is an undesired feature that is particularly present in low-cost and high-efficiency components \cite{nikos_prec}. When the input signal to the amplifier exhibits large amplitude variations, the non-linearities in the PAs may create severe spatial distortions. 

\subsection{Relevant Prior Art}

The most relevant multi-antenna works are \cite{oob_clarified,mollen_spatial,Salman2023a}. When it comes to MIMO arrays, only uniform linear arrays (ULAs) have been analyzed operating in the far-field region.
Specifically, \cite{oob_clarified} showed that in single-user beamforming, nonlinear distortion is beamformed into the same direction as the desired signal, while in multi-user beamforming, the distortion is beamformed into many distinct directions different from those of the desired signals. Moreover, \cite{mollen_spatial} showed that, if the input signal to the amplifiers has a dominant beam, the distortion is beamformed in the same way as that beam. When a superposition of multiple beams is transmitted without anyone being dominant, the distortion is practically isotropic. The transmission over frequency-selective channels gives rise to similar phenomena \cite{Salman2023a}.
Other studies \cite{ove, fredrik} highlight how nonlinear distortion critically affects MIMO systems and the trade-offs it imposes between system capacity and power efficiency. These insights reinforce the need to develop theoretical models that analyze the joint spatial and frequency behavior of such impairments, as done in this work, so the performance trade-offs can be rigorously optimized.

\subsection{Contributions}

Although the behavioral modeling of distortion has been studied in the multi-antenna literature, to the best of our knowledge, its effects on ELAAs and active RISs operating in the near-field region have not been addressed before, except in the preliminary version \cite{eucap2024} of this manuscript, which studied only active RISs but operating in the far-field region. 
In this paper, we theoretically analyze the distortion introduced by an ELAA during transmission or by an active RIS during signal reflection. For both cases, the arrays are modeled as uniform planar arrays (UPAs), and their operation is examined in both the radiative near-field and far-field regions.
Our main contributions can be summarized as follows:
\begin{itemize}
    \item We analyze the nonlinear distortion radiated from a transmitter (i.e., either ELAA or active RIS) equipped with a UPA. This setup enables us to analyze the distortion in both the azimuth and elevation domains.
    \item We consider near-field compliant channel models that  enable us to analyze the nonlinear distortion in the radiative near-field, where the depth domain can be controlled. In combination with the assumption of a transmitter equipped with a UPA, it provides us with the 3D behavioral analysis of distortion in the entire space of the channel, that is, in the azimuth, elevation and depth domains.
    \item We derive a theorem that analytically characterizes the angular directions and depth of nonlinear distortion in the near-field region. This theoretical framework provides a foundation for developing signal processing strategies to mitigate distortion, thereby improving the reliability and performance of the communication system. To illustrate its practical utility, we present a scenario where the theorem is used to predict distortion beamforming directions and adjust user scheduling in frequency and space to reduce its impact.    
    \item We refine the number of nonlinear distortion directions that were derived in \cite{mollen_spatial} for the far-field. Specifically, it was stated in \cite[Th.~2]{mollen_spatial} that third-degree non-linear distortion is beamformed in $(K^3-K^2+2K)/2$ directions. However, we show that this number is an upper bound and the exact number of directions can be substantially lower than that. Our results hold for both the near-field and far-field regions.
\end{itemize}

Our analytical and numerical results reveal that when an ELAA (active RIS) transmits (receives) data signals to (from) multiple users, the transmitted (reflected) signal can be distorted by its PAs and its spatial characteristics depend on the position of the users. As an example, the azimuth position of the users does not affect the beamforming directions of distortion on the elevation domain. However, the position of users in the elevation domain can affect the directions of distortion in the azimuth domain. Finally, when the users are located at different near-field distances from the ELAA or RIS, then the distortion is focused at distinct distances which are different from those of the position of the users.

\subsection{Notation}

Lowercase and uppercase boldface letters denote column vectors and matrices, respectively. 
For a matrix $\matA$,  we denote its complex conjugate and transpose by $\matA^{\ast}$ and $\matA^{T}$, respectively. 
The cardinality of a set $\mathcal{S}$, it is denoted as $|\mathcal{S}|$ which is the number of members of $\mathcal{S}$. 
For a vector $\vecv\in \mathbb{C}^{N}$, we use $\diag(\vecv)$ to denote an $N\times N$ diagonal matrix with the elements of vector $\vecv$ being its main diagonal. Finally, we use $\|\cdot\|$ to denote the ${l}_2$-norm of a vector. 

\subsection{Paper Outline}

The rest of the paper is organized as follows. In Section II, we introduce the system model of ELAAs as well as the propagation model for the radiative near-field channel. In Section III, we provide the analytical approximations of distortion's angular directions and depth in Fresnel region as well as analyze the distortion's behaviour in ELAA systems. In Section IV, we revise the nonlinear distortion's analysis assuming an active RIS. In Section V, we provide numerical results and validate the accuracy of our approximations for different scenarios. The paper is concluded in Section VI.

 

\section{System Model}

We consider a single-cell MIMO-Orthogonal Frequency Division Multiplexing (OFDM) system where an ELAA serves $K$ single-antenna users. The ELAA consists of a UPA with $M$ elements, which are arranged in  the $yz$-plane with $M_y$ and $M_z$ elements on the horizontal ($y$-axis) and vertical ($z$-axis) axes, respectively, such that $M=M_y\cdot M_z$. At the ELAA, the frequency-domain precoded vector signal is mapped to the time domain by performing an inverse discrete Fourier transform (IDFT) at each antenna element before being passed to a power amplifier which amplifies the transmitted time-domain signal. 

More specifically, we denote the discrete-time input and output signals to the amplifiers at the $m$-th antenna element and time sample $n$ as ${x}_{m}[n]$ and ${y}_{m}[n]$, respectively, where $m$ is the  $m=m_yM_z + m_z$ for $m_y = 0,\ldots,M_y-1$ and $m_z = 0,\ldots,M_z-1$. By denoting the operation of the PA at the $m$-th element as $\mathcal{A}(\cdot)$\footnote{All PAs among the ELAA elements are assumed to have the same transfer function. This fact can be seen as a worst case in terms of coherent combining of distortion \cite{z3ro_prec}. Small manufacturing variations will have negligible impact on the end results \cite{mollen_LNA}.}, the amplified transmit signal is given by
\begin{align} \label{eq:amplification}
    {y}_{m}[n] =  \mathcal{A}\left({x}_{m}[n]\right). 
\end{align}
For later use, we introduce the $M\times 1$ vector notation of the signals before and after amplification at time sample  $n$ as ${\vecx}_n = \left[{x}_{1}[n],\ldots,{x}_{M}[n]\right]^T$ and ${\vecy}_n = \left[{y}_{1}[n],\ldots,{y}_{M}[n]\right]^T$, respectively.
The discrete baseband transfer function of the nonlinear PA is
modeled by a memory polynomial and its output is related to its input by 
\begin{align}\label{eq:non_linear_model}
    \mathcal{A}\left(x_m[n] \right) = \sum_{p=0}^P\sum_{l=0}^L \beta_{2p+1}[l]x_m[n-l]\left|x_m[n-l]\right|^{2p},
\end{align}
where $\{\beta_{2p+1}\} \in \mathbb{C}$ are the polynomial parameters for the PA.

\begin{figure} 
        \centering 
	\begin{overpic}[width=\columnwidth,tics=10]{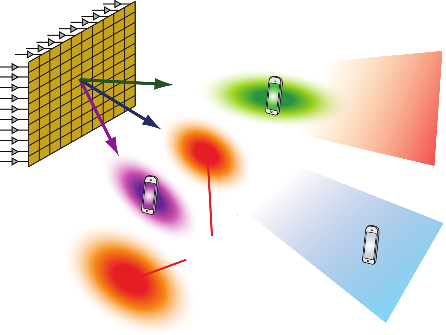}
   \put(59,58){\small User 1}
   \put(20.5,27){\small User 2}
   \put(79,11){\small User 3}
   \put(5,66){\small ELAA}
   \put(45,17){\small Near-field}
   \put(45,13){\small distortion}
   \put(84,52){\small Far-field}
   \put(84,48){\small distortion}
\end{overpic} \vspace{-3mm}
        \caption{The considered setup with an ELAA with $M$ antennas that transmit user signals using power amplifiers that create nonlinear distortion, resulting in distortion focused at different places in the near- and far-field.}
\label{figure_ELAA_setup} 
\end{figure}

The considered setup is illustrated in Fig.~\ref{figure_ELAA_setup} where an ELAA transmits to three users, two located in the near-field and one in the far-field. The non-linear distortion in the power amplifiers will give rise to distortion that is focused at specific locations in the near- and far-field, which will be characterized in this paper.

\begin{figure}[t!]
	\centering 
	\begin{overpic}[width=\columnwidth,tics=10]{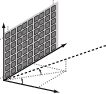}
		\put(54,1){\footnotesize $x$}
		\put(60,39){\footnotesize $y$}
		\put(9.5,57){\footnotesize $z$}
		\put(26.5,10){\footnotesize  $\varphi$}
		\put(40,17){\footnotesize  $\theta$}
		\put(5,11){\footnotesize  $1$}
		\put(1.5,41){\footnotesize  $N_{y}$}
		\put(12.7,54){\footnotesize  $1$}
		\put(46,74){\footnotesize  $N_{z}$}
		\put(80,31){\footnotesize  LoS path}
\end{overpic} 
	\caption{The 3D geometry of an RIS consisting of $M_{y}$ elements per row and $M_{z}$ elements per column.}\vspace{-0.3cm}
	\label{figure_geometric_setup}  
\end{figure}

\subsection{Near- and Far-field Channel Models}

We will now consider the radiative near-field and utilize the Fresnel region of the array response vector $\veca(\varphi,\theta, r)$ between a UPA of $M$ identical antennas and a point of interest $\boldsymbol{\psi} = (\varphi,\theta, r)$. Here, $\varphi$ and $\theta$ can specify either the incoming wave arrival direction from a user located at the point $\boldsymbol{\psi}$ or the outgoing wave direction from the UPA towards the point $\boldsymbol{\psi}$ in terms of azimuth and elevation angles, respectively. Additionally, the value $r$ is the distance between the first element of the UPA (i.e., $m=0$) and the point of interest $\boldsymbol{\psi}$. 
The array response vector $\veca(\boldsymbol{\psi}) \in \mathbb{C}^{M \times 1}$ can be expressed as
\begin{align}\label{eq:near_field_channel}
    \veca\left(\boldsymbol{\psi} \right) = \left[ 1, e^{j\frac{2\pi}{\lambda}(r-r_2)},\dots,e^{j\frac{2\pi}{\lambda}(r-r_M)} \right]^T,
\end{align}
where $r_m\ (n=1,\dots, M)$ denotes the distance between the $m$-th element of the UPA and the point $\boldsymbol{\psi}$, where $r = r_1$. Assuming that the first (that is, the reference) element of the UPA is placed at the origin of the Cartesian coordinate system, i.e., at $(0,0,0)$ and expressing the point of interest $\boldsymbol{\psi}$ in polar coordinates as $\boldsymbol{\psi} = (r\cos\varphi\cos\theta,r\sin\varphi\cos\theta,r\sin\theta)$, we have
\begin{align}
\begin{split}\label{eq:r_m}
        r_m = \left( r^2 + k^2_y(m) + k^2_z(m) - 2rk_z(m)\sin\theta \right. \\
        \left. - 2rk_y(m)\sin\varphi \cos\theta \vphantom{k^2_y(m)}\right)^{1/2},
\end{split}
\end{align}
where $k_z(m) \triangleq \mod(m,M_z)d_z$ and $k_y(m) \triangleq \lfloor \frac{m}{M_z} \rfloor d_y$ are the horizontal and vertical coordinates of the $m$-th element, and $\mod$ is the modulo operator. Moreover, $d_y$ and $d_z$ are the horizontal and vertical inter-element spacing, respectively.
For later use, we simplify the relative distance $r-r_m$ using the first-order Taylor approximation, i.e.,  $\sqrt{1-x} \approx 1 - x/2$ which yields
\begin{align}
    \phi_m(\varphi,\theta,r)  & \triangleq  r-r_m \label{eq:phase_def}\\
    &\approx \tilde{\phi}_m(\varphi,\theta,r)\notag\\
    &= k_z(m)\sin\theta  + k_y(m)\sin\varphi \cos\theta\label{eq:phase_apprx}\\
    &\hspace{22mm}- \frac{k^2_y(m) + k^2_z(m)}{2r}.\notag
\end{align}

The expression \eqref{eq:phase_apprx} is called Fresnel approximation and describes the spherical waves as parabolic \cite{Lozano2023}. We will use this approximation for the analytical derivations of distortion's beamformed directions, but we will use the exact expression in \eqref{eq:r_m} for numerical validation.

Furthermore, notice that the array response vector $\veca(\boldsymbol{\psi})$ in \eqref{eq:near_field_channel} can also be used when the point of interest  $\boldsymbol{\psi}$ is in the far-field region. In this case, the distance $r$ is considerably larger than the aperture of the UPA, i.e., $r\gg k^2_y(m) + k^2_z(m)$ and thus, the last term in \eqref{eq:phase_apprx} can be neglected. Therefore, we have $\phi_m(\varphi,\theta,r) \approx k_z(m)\sin\theta  + k_y(m)\sin\varphi \cos\theta$ and in this case the array response vector $\veca(\boldsymbol{\psi})$  becomes
\begin{align} \label{eq:array-response}
    \begin{split}
        \veca(\varphi,\theta) = \left[  1,\dots,e^{j\frac{2\pi}{\lambda}\left(
        k_{z}(m)\sin\theta +  k_{y}(m)\cos\theta \sin\varphi    \right)} \right. \\
         \left. ,\dots,  e^{j\frac{2\pi}{\lambda}\left(
        k_{z}(M)\sin\theta +  k_{y}(M)\cos\theta \sin\varphi    \right)} \right]^T,
    \end{split}
\end{align}
which is independent of the distance $r$ and the array response vector in the far-field region. 

Furthermore, the line-of-sight spherical wave channel model in the near-field region between the base station and user $k$ at the $\nu$-th subcarrier can be presented as\cite{Sherman1962,Dai2022,Symeon2023}
\begin{align}\label{eq:ofdm_channel_user}
    \hat\vech_k[\nu] = g_k\veca(\varphi_k,\theta_k,r_k) e^{-j2\pi \tau_k \nu},
\end{align}
where $g_k$ and $\tau_k$ are the complex path gain and delay of the signal from the first antenna at the base station to user $k$, respectively.
On this basis, the MIMO channel matrix $\hat{\matH}_\nu\in \mathbb{C}^{M\times K}$ at the $\nu$-th subcarrier can be expressed as
\begin{align}\label{eq:mimo-channel-freq}
       \hat{\matH}_\nu =  \underbrace{\left[ {\veca}_{1}\ {\veca}_{2}\ \dots\ {\veca}_{K}\right]}_{\triangleq \matA}\matF_\nu,
\end{align}
where ${\veca}_{k} \triangleq \veca(\varphi_k,\theta_k,r_k)$ is the channel (steering) vector between the base station and $k$-th user and the $K\times K$ matrix $\matF_\nu$ is diagonal with its $k$-th diagonal entry being $g_ke^{-j2\pi \tau_{k} \nu}$.

\subsection{Linear Precoded Signal}

At the ELAA, the data symbols for the $K$ users are mapped to the antenna array by a precoder. We denote the transmitted symbol on subcarrier $\nu$ to user $k\ (k=1,\ldots,K)$ as $s_{k}[\nu]$, while the corresponding $K\times 1$ data symbols vector is denoted as $\vecs_\nu = [s_{1}[\nu],\ldots,s_{K}[\nu]]^T$. 

We assume that the ELAA has access to perfect channel state information (CSI), i.e., it has perfect knowledge of the realizations of the frequency-domain channel matrices $\{{\hat\matH}_\nu\}$ for $\nu\in\mathcal{S}$. We make this assumption since the goal is to characterize the impact of nonlinear distortion. If the CSI is imperfect, the signals will be transmitted toward the wrong locations and the distortion will be determined by these believed user locations rather than the exact ones, but our results otherwise hold.

In general, the channel at a given subcarrier is frequency flat. Consequently, the data symbols are individually precoded for each subcarrier using the precoder $\matP_{\nu}=\mathcal{P}({\hat\matH}_{\nu})$, which is frequency flat and defined as a function $\mathcal{P}: \mathbb{C}^{K\times M}\rightarrow \mathbb{C}^{M\times K}$ of the channel. Specifically, the $M\times 1$ frequency-domain precoded symbol on subcarrier $\nu$ is expressed as $\hat\vecz_\nu=\matP_\nu\vecs_\nu$. 
The OFDM time-domain vector $\vecx_n$ at time sample $n$, before amplification, is obtained by applying the IDFT to the precoded frequency-domain symbols and is given by:
\begin{align}\label{eq:bb_tx_vector}
\vecx_n = \frac{1}{\sqrt{N}}\sum_{\nu\in \mathcal{S}}{\matP_\nu}\vecs_\nu e^{j\nu\frac{2\pi}{N}n},
\end{align}
 for $n=0,...,N-1$. In essence, the frequency-domain precoded symbol $\hat\vecz_\nu$ is transformed into the time domain via an IDFT.

Common linear precoders are the maximum-ratio transmission (MRT) and the zero-forcing (ZF), given by the following expressions:
\begin{align}
    \mathcal{P}_{\tt MRT}\left(\hat{\matH}_\nu\right) &=  \alpha {\matA^{\ast}}\matF_\nu^{\ast},\label{eq:mrt}\\
    \mathcal{P}_{\tt ZF}\left(\hat{\matH}_\nu\right) &=  \alpha \matA^{\ast}\underbrace{\left( \matF_\nu\matA^T \matA^{\ast} \right)^{-1}}_{\triangleq \matB_{\nu}},\label{eq:zf}
\end{align}
respectively, where the matrix $\matB_{\nu}\in \mathbb{C}^{K\times K}$ is defined for later use. Furthermore,  $\alpha$ is a constant used for power normalization that is chosen such that the transmitted signal before the amplification $\{\vecx_n\}$ satisfies the average transmit power constraint
\begin{align}
    \mathbb{E}\left[ \sum_{n=0}^{N-1}\|\vecx_n\|^2 \right] \le PS,
\end{align}
where $0<P<\infty$ is the average transmit power at the base station and $S\le N$ is the number of occupied subcarriers.

Note that in the expressions \eqref{eq:mrt} and \eqref{eq:zf},  we assume a multi-user (MU) scenario where all $K$ co-scheduled users occupy the same set of subcarriers.
With MRT, the resulting time-domain vector $\vecx_n$ simplifies to:
\begin{align}\label{eq:narrowband_tx_vector}
\vecx_n = {\matA^{\ast}}\vecs_n,
\end{align}
where $\vecs_n \triangleq ({1}/{\sqrt{N}})\sum_{\nu\in \mathcal{S}} \alpha \matF_{\nu}^{\ast}\vecs_\nu \exp\left\{ j2\pi \nu n/N \right\}$. It is important to note that when using the ZF, we obtain a similar expression to \eqref{eq:narrowband_tx_vector}, except that the vector $\vecs_n$ depends on $\matB_\nu$.

Let $e^{j\phi_m\left(\varphi_k,\theta_k,r_k \right)}$ represent the $(m,k)$-th entry of $\matA^{\ast}$, where $\phi_{m}\left(\varphi_k,\theta_k,r_k \right)\in \mathbb{R}$ is the phase of the precoder, respectively.
Then, from (\ref{eq:narrowband_tx_vector}), the transmitted symbol before amplification at the $m$-th antenna element can be expressed as 
\begin{align} \label{elaa:bb_tx_signal}
 x_m[n] = \sum_{k=1}^K e^{-j\frac{2\pi}{\lambda}\phi_{m}\left(\varphi_k,\theta_k,r_k \right)} s_k[n],
\end{align}
where $s_k[n]$ is the $k$-th entry of the vector $\vecs_n$. Notice that since we assume a line-of-sight (LoS) channel, the phase of the precoder $\phi_m(\cdot,\cdot,\cdot)$ is a conjugate match of the channel's phase. In this case, the transmitted signal $x_m[n]$ is focused at the points $(\varphi_k,\theta_k,r_k)$ for $k = 1,\dots, K$. 

\section{Nonlinear Distortion Analysis}

We are interested in the spatial characteristics of the nonlinear term (or distortion) in (\ref{eq:non_linear_model}) radiated from an ELAA. In the following theorem, we derive the Fresnel approximation for the azimuth angle, elevation angle, and range of the $p$-th order distortion in the near-field.

\begin{theorem} \label{th:nld_char}
Suppose the baseband signal in (\ref{elaa:bb_tx_signal}) is passed
through the non-linear function defined in (\ref{eq:non_linear_model}). In that case, the radiative near-field radiation pattern of the $p$-order non-linear distortion is beamformed in multiple directions $(\varphi_{\boldsymbol k},\theta_{\boldsymbol k},r_{\boldsymbol k})$ which can be approximated analytically as follows:  
    \begin{align}
     \theta_{\boldsymbol k} & \approx \arcsin\left( \sum_{i=0}^{2p} (-1)^{i} \sin\theta_{k_i}   \right),\label{eq:elevation} \\
    \varphi_{\boldsymbol k}  & \approx \arcsin\left( \frac{1}{\cos\theta_{\boldsymbol k}} \sum_{i=0}^{2p} (-1)^{i} \sin\varphi_{k_i}\cos\theta_{k_i} \right), \label{eq:azimuth} \\
    r_{\boldsymbol k} & \approx \left( \sum_{i=0}^{2p} (-1)^{i} \frac{1}{r_{k_i}} \right)^{-1}.\label{eq:range} 
    \end{align}
    where ${\boldsymbol k} \triangleq \left( k_{0},k_{1},\dots, k_{2p}\right)$ with ${\boldsymbol k}\in \{1,2,\ldots,K\}^{2p}$, $\varphi_{\boldsymbol k}\in [-\frac{\pi}{2}, \frac{\pi}{2}]$, $\theta_{\boldsymbol k}\in [-\frac{\pi}{2}, \frac{\pi}{2}]$ and $r_{\boldsymbol k}\ge 0$ are the azimuth, elevation and range of each focal point, respectively. 
    \begin{IEEEproof}
    See Appendix \ref{appendix:theorem1}.
    \end{IEEEproof}
\end{theorem}

The error of the Fresnel approximation used in Theorem \ref{th:nld_char} is small when $-2\tilde{\phi}_m(\varphi, \theta, r)/r\le 0.1745$, which results in errors below $3.5\cdot 10^{-3}$\cite{emil_pwr_scale},\cite[cf. expression (12)]{alva}. In practice, this can be achieved when the distance between the ELAA and a user is approximately at least 6 meters. 

\begin{remark}
The derivation of Theorem 1 assumes that all users allocate the same set of subcarriers. However, the results of the theorem also hold in the case of sub-band scheduling, where different users allocate distinct portions of the bandwidth. In this scenario, the transmit signal for each user can still be expressed in the form shown in \eqref{eq:narrowband_tx_vector}, with appropriate modifications to $\vecs_n$ to account for the allocation of different subcarriers. This ensures that the key principles and conclusions derived in the theorem remain valid under sub-band scheduling.
\end{remark}

Theorem \ref{th:nld_char} can also be generalized when $K_{\tt FF}$ $(0<K_{\tt FF}\le K)$ of the served users are in the far-field. With the next corollary we consider the extreme case where all users are in the far-field region.
\begin{corollary}\label{cor:farfield}
    Assuming that all users are in the far-field region $(K_{\tt FF}=K)$, i.e., $r_{k}\rightarrow \infty$ $\forall k$, the elevation and azimuth directions of the radiated distortion are given by the expressions \eqref{eq:elevation} and \eqref{eq:azimuth}, respectively. The expression for the $r_{pqv}$ in \eqref{eq:range} is ignored since it does not have physical meaning for far-field channels.
\end{corollary}

In the more general case where $K_{\tt FF}<K$ users are in the far-field, then as Corollary \ref{cor:farfield} implies the directions of elevation and azimuth in expressions \eqref{eq:elevation} and \eqref{eq:azimuth}, respectively, are valid regardless of whether a user is in the far- or near-field region. On the other hand, the expression for the near-field range $r_{pqv}$ in \eqref{eq:range} is modified depending on which users are in the far-field region. To illustrate this, let $\mathcal{K}_{\tt FF}$ be the set of users that are in the far-field region. Then, $\forall k\in \mathcal{K}_{\tt FF}$ the corresponding term $1/r_k\ (k=1,\dots, K)$ in \eqref{eq:range} is ignored.

Moreover, notice that the focal points $(\varphi_{\boldsymbol k}, \theta_{\boldsymbol k},r_{\boldsymbol k})$  given in Theorem \ref{th:nld_char} are not unique. For instance, it is easy to see that for the third-order non-linear distortion $(\varphi_{ppq}, \theta_{ppq},r_{ppq})=(\varphi_{vvq},\theta_{vvq},r_{vvq})$, $\forall p,q,v$. Thus, with the next corollary, we provide all the unique focal points derived in Theorem \ref{th:nld_char}. 
\begin{corollary}\label{prop:uniqeness}
    Let $\mathcal{P}$ be the set of all unique focal points of the third-order non-linear distortion derived in Theorem \ref{th:nld_char}. Then, $\mathcal{P}$ is given as the union $\mathcal{P} = \mathcal{P}_1 \cup \mathcal{P}_2 \cup \mathcal{P}_3$ of the following three disjoint subsets
    \begin{align}
    \begin{split}
       \mathcal{P}_1 = \{(\varphi_{p},\theta_{p},r_{p}): p=1,\ldots,K\}, 
    \end{split}
    \\[1ex]
    \begin{split}
        \mathcal{P}_2 = \{(\varphi_{pqp},\theta_{pqp},r_{pqp}): p,q=1,\ldots,K,\\
        q\neq p, K\ge 2\},    
    \end{split}
    \\[1ex]
    \begin{split}
        \mathcal{P}_3 = \{(\varphi_{pqv},\theta_{pqv},r_{pqv}): p,q,v=1,\ldots,K, \\
        q\neq p, v\neq p,v>q, K\ge 3\}.
    \end{split}
    \end{align}
    \begin{IEEEproof}
        See Appendix B.
    \end{IEEEproof}
\end{corollary}

Notice that the $p$-th focal point in $\mathcal{P}_1$ represents the distortion beamformed at the location of the $p$-th user, where, as expected, $|\mathcal{P}_1| = K$. It can be observed from \eqref{eq:array_factor_cor} that these focal points carry more power than those in $\mathcal{P}_2$ and $\mathcal{P}_3$, which implies that distortion is expected to be stronger in the direction of the users.
Furthermore, it is worth mentioning that when the argument of the function $\arcsin(\cdot)$ in \eqref{eq:elevation} and \eqref{eq:azimuth} belongs to the subset of real numbers: $\left(-\infty,-1\right)\cup\left(1,\infty\right)$, then its corresponding value $(\theta_{pqv}$ or $\varphi_{pqv}$) is a complex number. In this case, the direction lacks a physical meaning and thus, it can be ignored since there is no physical location where the distortion can arise. Similarly, we consider only non-negative values of the range $r_{pqv}$. Therefore, the number of distortion directions of $\mathcal{P}_2$ and $\mathcal{P}_3$ in Corollary \ref{prop:uniqeness} are at most $|\mathcal{P}_2| \le K(K-1)$ and $|\mathcal{P}_3| \le K(K-1)(K-2)/2$ which implies that the total number of unique focal points are at most $|\mathcal{P}| \le (K^3-K^2+2K)/2$. This is in contrast to \cite[Th.~2]{mollen_spatial}, where it is stated that the third-degree distortion is beamformed exactly into $(K^3-K^2+2K)/2$ directions.

Theorem \ref{th:nld_char} brings important insights into the characteristics of nonlinear distortion. First, notice that the distance (or range) of users from the ELAA does not affect the distortion on the elevation or azimuth angles. For example, when an ELAA performs finite-depth beamforming, i.e., all co-scheduled users are in the same direction but different distances from the ELAA, i.e., at $(\varphi,\theta,r_k)$, $\forall k$, the distortion is beamformed towards the same direction but it is spread into distinct distances different from those of users according to the expression (\ref{eq:range}). 
Therefore, the distortion can have higher performance degradation in finite-depth beamforming than full-dimensional beamforming. The reason is that in finite-depth beamforming the distortion's radiation power is limited towards only one direction, while in full-dimensional beamforming, it is spread over the whole space resulting in lower power levels per directions.  

With the next corollaries, we highlight two special cases of how the third-order distortion is radiated in the azimuth and elevation domain in the $yz$-plane.

\begin{corollary}\label{cor:same_elevation}
Suppose all users are located in the same elevation direction (i.e., $\theta_k=\theta_o$, $\forall k$) but different azimuth angles. The third-order distortion is then focused at the points $(\varphi_{pqv},\theta_{pqv},r_{pqv})$:
\begin{align}
    \theta_{pqv}  &\approx \theta_o,\\
    \varphi_{pqv} &\approx \arcsin\left(\sin\varphi_p - \sin\varphi_q + \sin\varphi_v\right),\label{cor:azimuth}
\end{align}
and the distance $r_{pqv}$ is given in the expression \eqref{eq:range}.
\end{corollary}
According to Corollary \ref{cor:same_elevation}, the distortion is beamformed into the same elevation angle as that of the desired signal from the ELAA but in distinctly different directions in the azimuth domain. Furthermore, the azimuth directions are independent of the elevation angles of the users.

\begin{corollary}\label{cor:same_azimuth}
Suppose all users are located in the same azimuth direction (i.e., $\varphi_k=\varphi_o$, $\forall k$) but different elevation angles. Then, the third-order distortion is beamformed into the directions $(\varphi_{pqv},\theta_{pqv},r_{pqv})$:
\begin{align}
    \theta_{pqv}  &\approx \arcsin\left(\sin\theta_p - \sin\theta_q + \sin\theta_v \right),\label{cor:elevation}\\
    \varphi_{pqv} &\approx \arcsin\left(\frac{\sin\varphi_o}{\cos\theta_{pqv}} \left(  \cos\theta_p - \cos\theta_q + \cos\theta_v \right) \right),
\end{align}
and the distance $r_{pqv}$ is given in the expression \eqref{eq:range}.
\end{corollary}

According to Corollary \ref{cor:same_azimuth}, the ELAA can create distortion which is beamformed into distinct directions different from those of the desired signals, and these are spread over both the azimuth and elevation domains. In particular, the azimuth direction is a function of the elevation implying that at different elevation angles, the distortion is beamformed towards different azimuth directions. Only when $\varphi_o = 0^{\circ}$, the distortion in the azimuth domain becomes independent of the elevation angles and it is is beamformed into the zero angle.

Finally, it is worth mentioning that in the special case where the elevation angle $\theta_o$ in Corollary \ref{cor:same_elevation} and the azimuth angle $\varphi_o$ in Corollary \ref{cor:same_azimuth} are zero, then the distortion's directions in \eqref{cor:azimuth} and \eqref{cor:elevation}, respectively, become identical with those of a ULA and therefore these corollaries coincide with \cite[Th.~2]{mollen_spatial}.


\section{Nonlinear Distortion of an Active RIS}

\begin{figure} 
        \centering 
	\begin{overpic}[width=\columnwidth,tics=10]{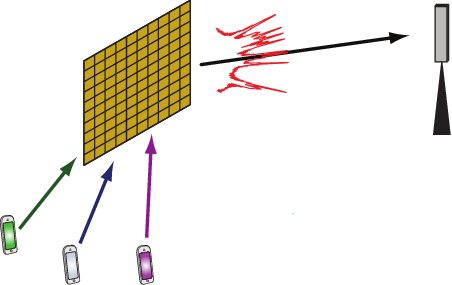}
  \put(36,5){\small $K$ transmitting users}
   \put(67,29){\small Receiving base station}
   \put(15,58){\small Active RIS}
   \put(45,38){\small Reflected distorted signal}
\end{overpic} \vspace{-3mm}
        \caption{The considered setup with an active RIS with $M$ cells that reflects user signals toward a base station while creating nonlinear distortion.}
\label{figure_RIS_setup} 
\end{figure}

We now consider an active RIS with {$M$} reconfigurable {cells} with the same geometric array structure as in the previous sections. It is deployed to beamform the impinging signal from $K$ single-antenna users to a base station.
At the RIS, the time-domain signal received from the users at each reflecting {cell} is phase-shifted and then amplified before being re-radiated.\footnote{The order of phase-shifting and amplification has no impact on our results.} These amplifiers are nonlinear and this results in a reflected signal that is beamformed towards more locations than there are users, as illustrated in Fig.~\ref{figure_RIS_setup}.

\subsection{Phase-shifted Signal}

We let the vector $\vecs_n \in \mathbb{C}^{K\times 1}$ contain the transmitted time-domain OFDM symbols from the $K$ users and the matrix $\matA \in \mathbb{C}^{M\times K}$ be the channel between the RIS and users as defined in \eqref{eq:mimo-channel-freq}. The impinging signal at the RIS will then be $\matA\vecs_n$, while the phase-shifted signal $\vecx_n$ at time sample $n$ can be expressed as
\begin{align}\label{eq:signal_x}
    \vecx_n &= {\boldsymbol{\Phi}\matA\vecs_n},
\end{align}
where the diagonal matrix $\boldsymbol{\Phi}$ with unit-modulus entries contains the RIS phase shift configuration. 
We will denote the phase-shift in the $m$-th entry in \eqref{eq:near_field_channel} as 
\begin{align}
  \frac{2 \pi}{\lambda} \phi_m(\varphi,\theta,r),
\end{align}
and assume that the RIS is configured as
\begin{equation}
    \boldsymbol{\Phi} = \mathrm{diag}\left(e^{-j\frac{2 \pi}{\lambda}\phi_{1}(\varphi_s,\theta_s,r_{\tt s})},\ldots,e^{-j\frac{2 \pi}{\lambda}\phi_{M}(\varphi_{\tt s},\theta_{\tt s},r_{\tt s})}\right),
\end{equation}
to compensate for the phase shifts of a signal arriving from user $k$ in the direction
$(\varphi_{k},\theta_{ k},r_{k})$.
The phase-shifted signal in (\ref{eq:signal_x}) becomes
\begin{equation}
        \vecx_n = \boldsymbol{\Phi} \sum_{k=1}^K \veca\left( \varphi_{k}, \theta_{k},r_{\tt k} \right) s_k[n],
\end{equation}
where $s_k[n]$ is the $k$-th entry of the vector $\vecs_n$ and is the time-domain symbol transmitted by the $k$-th user.
Hence, the phase-shifted signal at the $m$-th \blue{cell} before amplification is
\begin{align}\label{eq:bb_tx_signal}
    {x}_m[n] = \sum_{k=1}^K e^{j\frac{2 \pi}{\lambda}\left( 
    {\phi}_m(\varphi_{ k},\theta_{k},r_{\tt k}) - 
    \phi_m(\varphi_{\tt s},\theta_{\tt s},r_{\tt s})
    \right)}s_k[n],
\end{align}
where $\phi_m(\varphi,\theta,r)$ is defined in \eqref{eq:phase_def}.
We notice in \eqref{eq:bb_tx_signal} that the effective phase of the $k$-th signal after applying the RIS phase shifting is $\phi_m (\hat{\varphi}_k,\hat{\theta}_k,\hat{r}_k) = {\phi}_m(\varphi_{ k},\theta_{k},r_k) - \phi_m(\varphi_{\tt s},\theta_{\tt s},r_{\tt s})$, where the effective azimuth angle $\hat{\varphi}_k$, elevation angle $\hat{\theta}_k$ and range $\hat{r}_{k}$ can be approximated analytically 
\begin{align}
  \begin{split}
        \hat\theta_k \approx \arcsin\left(\sin\theta_k - \sin\theta_{\tt s} \right) ,
    \end{split} 
       \\[1ex]
    \begin{split}
    \hat\varphi_{k} \approx \arcsin\left(\frac{1}{\cos\hat\theta_{k}} \left(  \cos\theta_k\sin\varphi_k  - \cos\theta_{\tt s}\sin\varphi_{\tt s} \right) \vphantom{\frac{1}{\cos\theta_{pqv}}}\right),
    \end{split}
     \\[1ex]
    \begin{split} 
        \hat{r}_{k}  \approx \frac{r_{\tt s}\ r_k}{r_{\tt s} - r_k},\ r_{\tt s} \neq r_k
    \end{split}
\end{align}
by using the Fresnel approximation of $\phi_m(\cdot,\cdot,\cdot)$ in \eqref{eq:phase_apprx}. Notice from \eqref{eq:phase_apprx}, that in the case of $r_{\tt s} = r_k$, the phase of the re-directed signal, i.e., ${\phi}_m(\varphi_{ k},\theta_{k},r_k) - \phi_m(\varphi_{\tt s},\theta_{\tt s},r_{\tt s})$ is independent of the distances $r_{\tt s}$ and $r_k$ implying that the re-transmitted signal does not have a focal point in the near-field region but in far-field.

Without the RIS phase-shifts (i.e., if they are zero), the $k$-th signal arrives from $(\varphi_k,\theta_k)$ and would be reflected (mainly) towards the direction
$(-\varphi_k,-\theta_k)$. With the RIS phase-shifts, the signal is instead reflected in the direction $(-\hat\varphi_k,-\hat\theta_k)$.

\begin{figure*}
\begin{multicols}{2}
  \includegraphics[width=\linewidth]{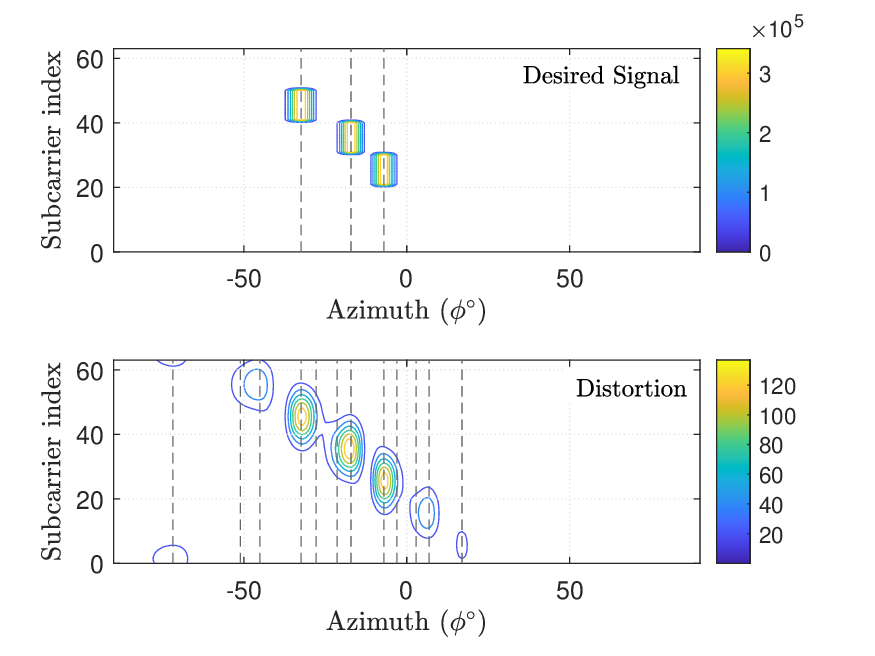}\caption*{(a) Elevation plane at $-\hat\theta_o$}
  \includegraphics[width=\linewidth]{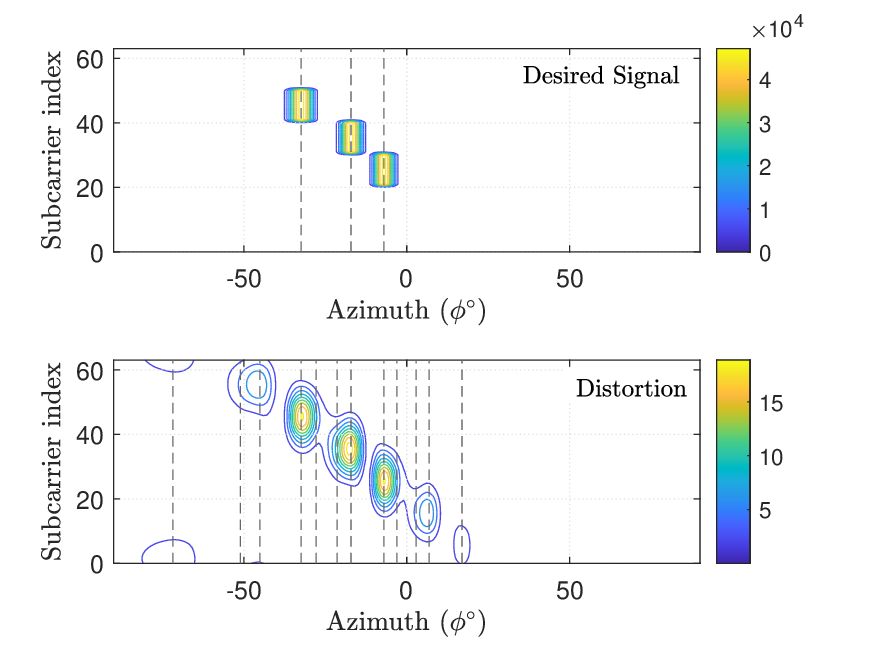}\caption*{(b) Elevation plane at $\theta_o$}
\end{multicols}  \vspace{-3mm}
\caption{The PSD [dBm/Hz] of the far-field linear signal (top) and nonlinear distortion (bottom) radiated from an actve RIS with $20\times 20$ antenna cells as a function of the subcarrier index and azimuth angle. All users are seen from the same elevation angle of $\theta_o = 0^{\circ}$. The vertical lines represent the theoretical azimuth directions of the distortion obtained in Theorem 1.} \label{figure:azimuthSimulation}
\end{figure*}

\subsection{Nonlinear Distortion}

With the next corollary, we provide the spatial characteristics of
the third-order nonlinear term (or distortion) radiated
from an active RIS.

\begin{corollary}\label{ris_corollary}
Suppose the baseband signal in (\ref{eq:bb_tx_signal}) is passed
through the non-linear function in (\ref{eq:non_linear_model}). Then, the radiation pattern of the $p$-th order non-linear distortion is beamformed
in a multitude of directions $(-\varphi_{\boldsymbol k},-\theta_{\boldsymbol k},r_{\boldsymbol k})$  approximated analytically by      
\begin{align}
     \theta_{\boldsymbol k} & \approx \arcsin\left( \sum_{i=0}^{2p} (-1)^{i} \sin\hat\theta_{k_i}   \right),\label{ris:elevation} \\
    \varphi_{\boldsymbol k}  & \approx \arcsin\left( \frac{1}{\cos\theta_{\boldsymbol k}} \sum_{i=0}^{2p} (-1)^{i} \sin\hat\varphi_{k_i}\cos\hat\theta_{k_i} \right), \label{ris:azimuth} \\
    r_{\boldsymbol k} & \approx \left( \sum_{i=0}^{2p} (-1)^{i} \frac{1}{\hat{r}_{k_i}} \right)^{-1}.\label{ris:range} 
    \end{align}
    where ${\boldsymbol k} \triangleq \left( k_{0},k_{1},\dots, k_{2p}\right)$ with ${\boldsymbol k}\in \{1,2,\ldots,K\}^{2p}$, $\varphi_{\boldsymbol k}\in [-\frac{\pi}{2}, \frac{\pi}{2}]$, $\theta_{\boldsymbol k}\in [-\frac{\pi}{2}, \frac{\pi}{2}]$ and $r_{\boldsymbol k}\ge 0$ are the azimuth, elevation and range of each focal point, respectively. 
\end{corollary}

Notice that Corollary \ref{ris_corollary} is identical with Theorem \ref{th:nld_char}. Therefore, the behavioural analysis derived in Theorem \ref{th:nld_char} for an ELAA system, it can be also applied for an active RIS. However, in the latter case, the beamformed vector of each signal is not created in the baseband but comes from the channel from the users as well as we can modify the reflected angles using the RIS. 

\section{Numerical Results}

\begin{figure*}
\begin{multicols}{2}
  \includegraphics[width=\linewidth]{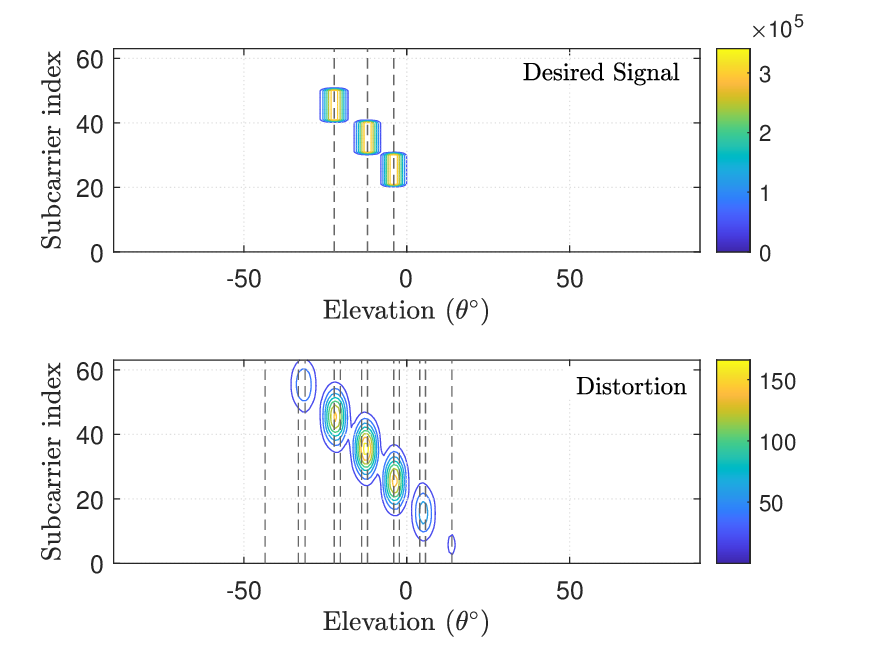}\caption*{(a) Azimuth plane at $-\hat{\varphi}_o$}
  \includegraphics[width=\linewidth]{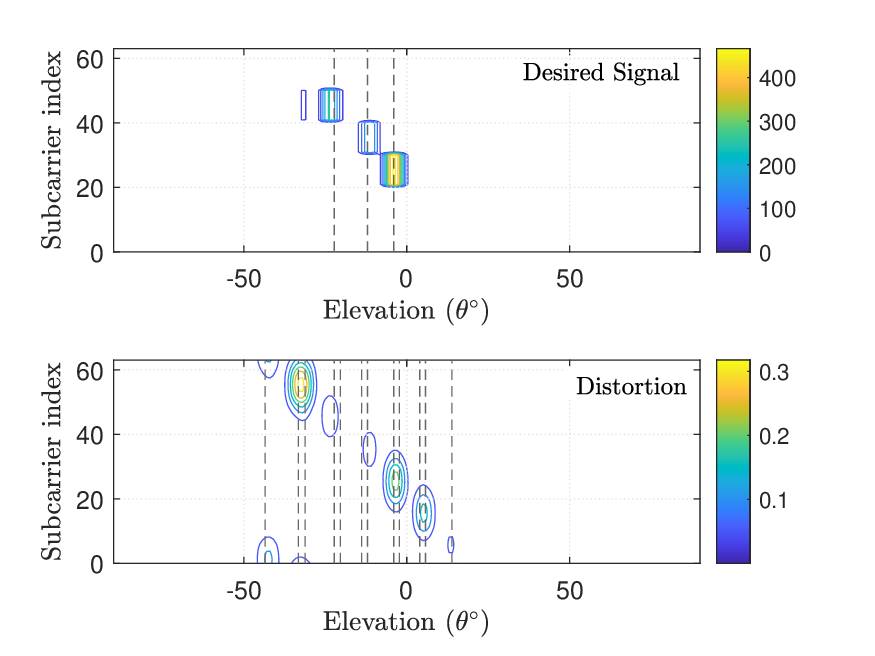}\caption*{(b) Azimuth plane at $\varphi_o$}
\end{multicols} \vspace{-3mm}
\caption{The PSD [dBm/Hz] of the far-field linear signal (top) and nonlinear distortion (bottom) radiated from an active RIS with $20\times 20$ cells as a function of the subcarrier index and elevation angle. All users are seen from the same azimuth angle of $\varphi_o = 14^{\circ}$. The vertical lines represent the theoretical elevation directions of the distortion obtained in Theorem 1.} \label{figure:elevationSimulation}
\end{figure*}

\begin{figure*}
\begin{multicols}{2}
  \includegraphics[width=\linewidth]{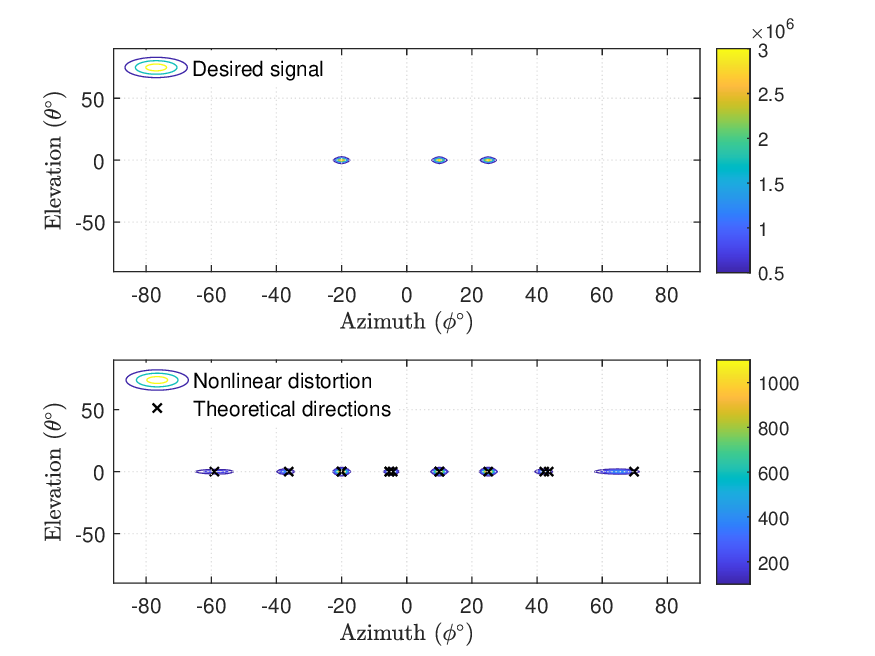}\caption*{(a) Azimuth beamforming}
    \includegraphics[width=\linewidth]{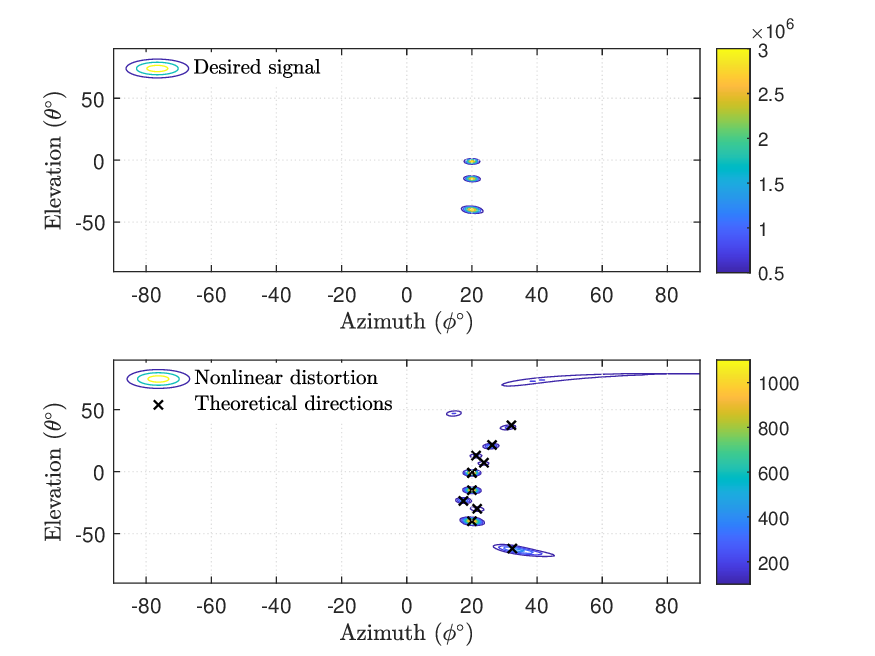}\caption*{(b) Elevation beamforming}
\end{multicols}
\caption{The PSD [dBm/Hz] of the radiative near-field linear signal (top) and nonlinear distortion (bottom) for three different cases of users position. In this example, we assume a UPA $35\times 35$, carrier frequency $f=3$ GHz. In all cases, all users are located at the range of 20m from the ELAA.}\label{figure:2dBeamformingSimulationElaa}
\end{figure*}

In this section, we present simulation results to validate our theoretical analysis and the tightness of the approximations. We consider that both ELAA and RIS are equipped with a UPA with half-wavelength antenna spacing. 
For simplicity, we assume that the nonlinearities can be modeled by a third-order nonlinear PA model, given by:
\begin{align} \label{eq:third-order_pa}
    \mathcal{A}(x_m[n]) = \beta_1 x_m[n] + \beta_3 x_m[n] \lvert x_m[n] \rvert^2,
\end{align}
where, unless otherwise specified, the parameters are selected such that the error vector magnitude (EVM) at the output is $3\%$. Specifically, $\beta_1 = 1.042$ and $\beta_3 = -0.0212$. 
Moreover, we assume that the RIS phase shifter rotates the impinging signal by $\left( \varphi_s, \theta_s \right) = (-2^{\circ}, -4^{\circ})$ to target the base station.

\subsection{Near- and Far-Field Distances for Antenna Arrays}

The near-field can be divided into reactive and radiative, where the latter is also known as Fresnel region. The radiative near-field starts at the Fresnel distance\cite{emil_pwr_scale}. Furthermore, the radiative near-field and far-field are separated by the Fraunhofer array distance $d_{\rm FA} = \frac{2 \Delta^2}{\lambda}$, where $\Delta$ is the maximum linear dimension of the array and $\lambda$ is the wavelength. We are interested in the radiative near-field region where there exists only phase variation over the array, characterized by spherical wavefronts. This region occurs when the propagation distance is larger than $d_{\rm B} = 2 \Delta$ but smaller than $d_{\rm FA}$\cite{emil_pwr_scale,alva}. 

In this section, we assume that the carrier frequency is $3$GHz, the antenna spacing is $\lambda/2$ and the UPA has the same number of elements per dimension, i.e., $M_z=M_y=M_o$.  Therefore, its maximum linear dimension is $\Delta = \lambda (M_o-1)/\sqrt{2}$, where $\lambda=0.1$ meters. In the case, we have a UPA with $20\times 20$ elements, the radiative near-field starts at $d_B = 2.68$ meters and it ends at the propagation distance $d_{\rm FA} = 36.1$ meters. When the propagation distance is larger than $36.1$ m, the wavefront can be characterized by a planar wavefront, as observed in the far-field model.
When the size of the UPA increases to $35\times 35$ elements, then  the radiative near-field starts at $d_B = 4.8$ meters and it ends at the propagation distance $d_{\rm FA} = 115.6$ meters.

\subsection{Radiation Pattern}

To demonstrate the radiation patterns of the desired (linear) signal and nonlinear distortion, we use the Bussgang decomposition\cite{Demir2021a} to express the third-order memoryless PA output as
\begin{align}  \label{eq:bussgang}
    \vecy_n = \matG\vecx_n + \vecd_n,
\end{align}
where the scaled desired signal $\vecu_n\triangleq \matG\vecx_n$ is the linear part of the output while the nonlinear distortion is represented by $\vecd_n\in \mathbb{C}^{M \times 1}$, which is uncorrelated with $\vecx_n$.

The Bussgang gain matrix $\mathbf{G} = \mathbf{I}_M + 2\beta_3 \operatorname{diag}(\mathbf{C}_{\mathbf{x}\mathbf{x}}[0])$ is a diagonal matrix, where $\mathbf{C}_{\mathbf{x}\mathbf{x}}[\tau] = \mathbb{E}\{\mathbf{x}_n \mathbf{x}_n^{\mathrm{H}} \}$ is the covariance matrix of $\mathbf{x}_n$. Similarly, the covariance matrix of the distortion vector is given by $\mathbf{C}_{\mathbf{d}\mathbf{d}}(\tau) = \mathbf{C}_{\mathbf{y}\mathbf{y}}[\tau] - \mathbf{C}_{\mathbf{u}\mathbf{u}}[\tau]$, where $\mathbf{C}_{\mathbf{u}\mathbf{u}}[\tau] = \mathbf{G}\mathbf{C}_{\mathbf{x}\mathbf{x}}[\tau]\mathbf{G}$ and the output covariance $\mathbf{C}_{\mathbf{y}\mathbf{y}}[\tau]$ is given analytically in \cite[Eq. (27)]{mollen_icc}.
The radiation pattern in the frequency domain is given by the power spectral density (PSD) $\matS_{\bf{yy}}(f)$, which is found by taking the Fourier transform of $\matC_{\bf{yy}}[\tau]$. Then, the power radiated in the direction of $(\varphi, \theta)$ at frequency $f$ is obtained by multiplying by the corresponding array response vector:
\begin{align}\label{psd}
   \matS_{\varphi, \theta,r}(f) &= \veca^T\left( \varphi, \theta,r\right) \matS_{\bf{yy}}(f) \veca^{\ast}\left( \varphi, \theta,r\right).
\end{align}
To distinguish the desired linear signal from the undesired nonlinear distortion, we use the decomposition in (\ref{eq:bussgang}). Since the useful signal and distortion terms are uncorrelated, the PSD of the transmitted signal can be expressed as
\begin{align*}
  \matS_{\bf{yy}}(f) = \matS_{\bf{uu}}(f) + \matS_{\bf{dd}}(f),
\end{align*}
where $\matS_{\bf{uu}}(f)$ and $\matS_{\bf{dd}}(f)$ can be computed by taking the Fourier transforms of $\matC_{\bf{uu}}[\tau]$ and $\matC_{\bf{dd}}[\tau]$, respectively.
The radiation patterns of the in-band desired signal and in-band/out-of-band nonlinear distortion can be computed as $\veca^T\left( \varphi, \theta,r\right) \matS_{\bf{uu}}(f) \veca^{\ast}\left( \varphi, \theta,r\right)$ and $\veca^T\left( \varphi, \theta,r\right) \matS_{\bf{dd}}(f) \veca^{\ast}\left( \varphi, \theta,r\right)$, respectively, by following the same principle as in (\ref{psd}).

\begin{figure*}
\begin{multicols}{2}
\includegraphics[width=\linewidth]{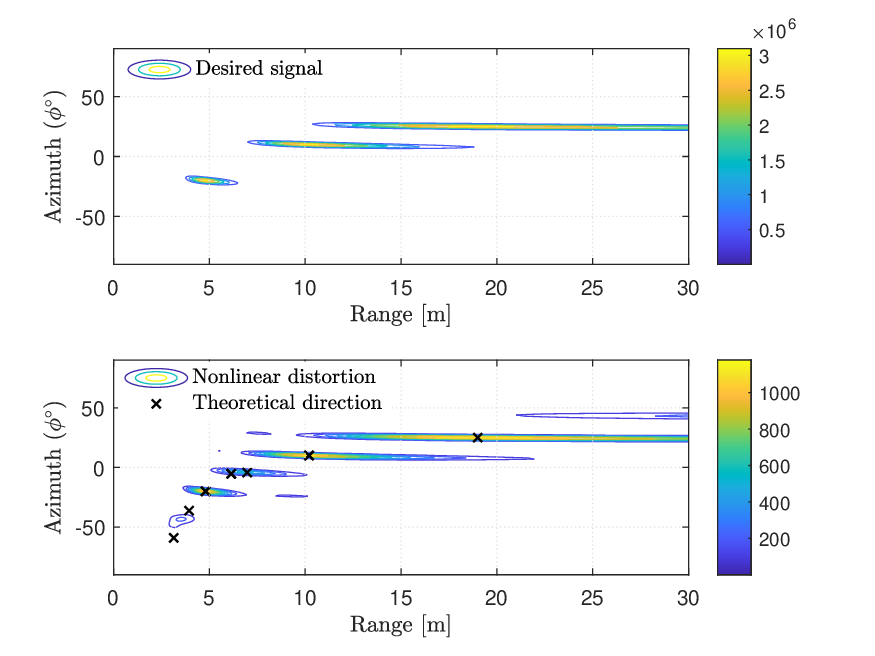}\caption*{(a)  Azimuth-depth beamforming}
 \includegraphics[width=\linewidth]{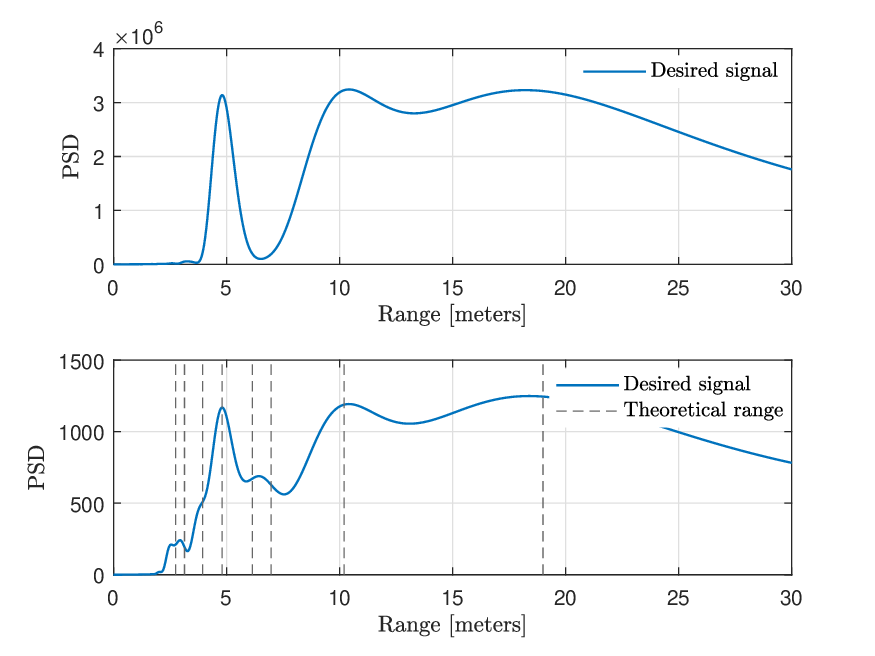}\caption*{(b) Finite-depth beamforming}
\end{multicols}
\caption{The PSD [dBm/Hz] of the radiative near-field linear signal (top) and nonlinear distortion (bottom) radiated from an ELAA. In this example, we assume a UPA 35 × 35 and carrier frequency f = 3 GHz.}\label{figure:fullDimensionalSimulated}
\end{figure*}

\begin{figure*}
\begin{multicols}{2}
  \includegraphics[width=\linewidth]{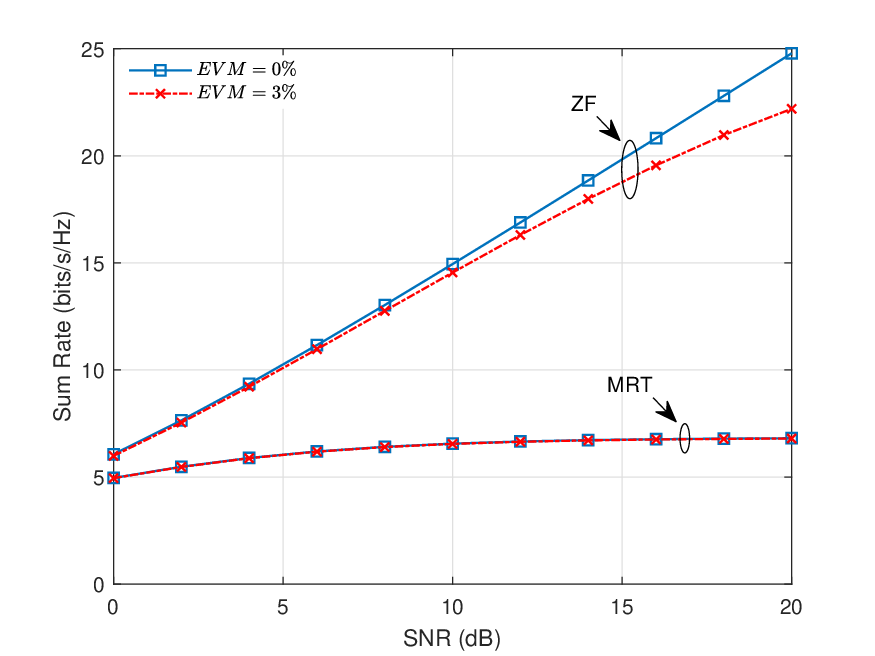}\par\caption{Sum rate as a function of SNR for different levels of EVM with $M=16$ antennas.}\label{fig:sum_rate_16}
    \includegraphics[width=\linewidth]{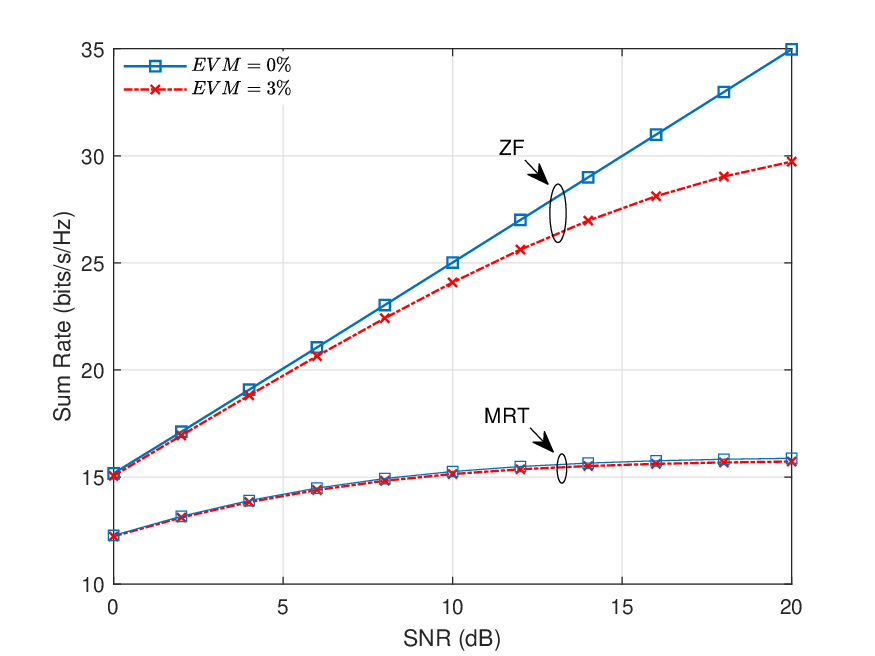}\par\caption{Sum rate as a function of SNR for different levels of EVM with $M=100$ antennas.}\label{fig:sum_rate_100}
\end{multicols}
\end{figure*}

\subsection{Distortion in the Frequency-Azimuth Domain}

In Fig.~\ref{figure:azimuthSimulation}, we show the PSD of the radiated signal from the RIS when it receives far-field signals from three users that are seen from the same elevation angle ($\theta_o=0^{\circ}$) but in different azimuth angles: $(\varphi_1,\varphi_2,\varphi_3) = (2^{\circ},20^{\circ},35^{\circ})$. Each user allocates a different set of subcarriers as shown in Fig.~\ref{figure:azimuthSimulation}, and might be served by the same base station or a different one. 

In Fig.~\ref{figure:azimuthSimulation}(a), we show the PSD of the desired signal (top) and nonlinear distortion (bottom) as a function of the subcarrier index and azimuth angle. Since the radiated field is three-dimensional, we consider the elevation plane that the RIS focuses the signals towards (i.e.,  $\hat\theta_o$). We notice that the RIS creates both in-band and out-of-band distortion that is beamformed into distinct directions; there are more directions than the number of users and distortion appears in bands where the desired signal does not exist. Nevertheless, most of the distortion power is beamformed in the same directions as the desired signal. Additionally, in Fig.~\ref{figure:azimuthSimulation}(b), we show the same functions but consider the original elevation plane where the users reside (i.e., $\theta_o$). As expected, we notice that the distortion is stronger towards the direction of the re-radiated signal, where the RIS signal is beamformed, than in the original plane. Moreover, we notice that as long as all users are located at the same elevation angle, the azimuth beamforming directions of the distortion are not affected by which elevation plane we consider. This is a consequence of Corollary \ref{cor:same_elevation}, which proves that the distortion radiated from the RIS is beamformed only in the azimuth domain. Finally, we notice that the theoretical distortion directions derived in Corollary \ref{cor:same_elevation} (shown as vertical lines) coincide with those obtained by the simulations.

\subsection{Distortion in the Frequency-Elevation Domain}

Similarly, in Fig.~\ref{figure:elevationSimulation}, we show the PSD of the radiated signal from the RIS when it receives far-field signals from three users seen from the same azimuth angle (i.e., $\varphi_o=14^{\circ}$) but in different elevation angles $(\theta_1,\theta_2,\theta_3) = (1^{\circ},8^{\circ},18^{\circ})$. 

In Fig.~\ref{figure:elevationSimulation}(a), we show the PSD of the desired signal (top) and nonlinear distortion (bottom) as a function of the subcarriers index and elevation angle in the azimuth plane that the re-radiated signal is aimed towards using the RIS. By contrast, in Fig.~\ref{figure:elevationSimulation}(b), we consider the original azimuth plane in the direction of the users. A main difference from Fig.~\ref{figure:azimuthSimulation} is that the elevation beamforming directions of the distortion depend on the azimuth direction. In other words, different azimuth directions yield different beamforming directions of the distortion in the elevation domain.
For example, we notice that in Fig.~\ref{figure:elevationSimulation}(a), some of the elevation beamforming directions of the distortion derived in Corollary \ref{cor:same_azimuth} are missing. The reason is that these elevation directions appear at different azimuth directions as shown in Fig.~\ref{figure:elevationSimulation}(b).


\subsection{Distortion in the Angular Domain}

Next, we illustrate Theorem \ref{th:nld_char} by showing the PSD of nonlinear distortion as a function of the azimuth and elevation planes. To facilitate this, all users occupy the same subcarrier indices and they are located at the same distance from the ELAA $(r_o=20\text{m})$. More precisely, in Fig.~\ref{figure:2dBeamformingSimulationElaa}(a), we show the PSD of the desired signal (top) and nonlinear distortion (bottom) radiated from an ELAA towards three co-scheduled users as a function of the elevation and azimuth angles. We assume two-dimensional azimuth beamforming, that is all users are located at the same elevation angle (i.e., $\theta_o=0^{\circ}$) but at different azimuth angles $(\varphi_1,\varphi_2,\varphi_3) = (-20^{\circ},10^{\circ},25^{\circ})$. 

As indicated in Corollary \ref{cor:same_elevation} the distortion is beamformed into the same elevation angle as that of the desired signals but at different azimuth directions. On the other hand, when we assume two-dimensional elevation beamforming, i.e., the users are located at the same azimuth (i.e., $\varphi_o=20^{\circ}$) but at different elevation angles $(\theta_1,\theta_2,\theta_3) = (-1^{\circ},-15^{\circ},-40^{\circ})$ as shown in Fig.~\ref{figure:2dBeamformingSimulationElaa}(b), the distortion will not only beamformed at different elevation angles than those of the desired signal but it will also spread at different azimuth angles than $(\varphi_o)$ as proved in Corollary \ref{cor:same_azimuth}.

\subsection{Distortion in the Depth Domain}

In Fig.~\ref{figure:fullDimensionalSimulated}(a), we assume that all three co-scheduled users occupy the same subcarrier indices and are located in the same elevation angle (i.e., $\theta_o = 0^{\circ}$). As it is shown in the the PSD of the desired signal (top), the distances to the users from the ELAA are $(r_1,r_2,r_3) = (4.8,9.8,19)$ and it is seen from the azimuth angle $(\varphi_1,\varphi_2,\varphi_3) = (-20^{\circ},10^{\circ},25^{\circ})$, respectively. First, notice that the nonlinear distortion  is focused at distinct points that are at different azimuth and distances from the ELAA than those of the desired signal as predicted in Theorem \ref{th:nld_char}. Additionally, notice that the azimuth directions of the users in this example are the same as those in the example of Fig.~\ref{figure:2dBeamformingSimulationElaa}(a). Thus, one would expect that the distortion's azimuth directions in these two examples are identical. However, in this example, due to the fact that the users are at different distances from the ELAA, the expression \eqref{eq:range} for the distortion's distances yields negative values for some azimuth directions which do not appear in the PSD of the Fig.~\ref{figure:fullDimensionalSimulated}(a). Therefore, the distortion's focal points in Fig.~\ref{figure:fullDimensionalSimulated}(a) are less than those in Fig.~\ref{figure:2dBeamformingSimulationElaa}(a).

Finally, in Fig.~\ref{figure:fullDimensionalSimulated}(b), we demonstrate the tightness of the approximate closed-form expression of focusing points of nonlinear distortion derived in Theorem \ref{th:nld_char}. We assume finite-depth beamforming, that is, all three users are located at the same azimuth and elevation direction but at different near-field distances from the ELAA: $(r_1,r_2,r_3)$. 
Notice that the distortion can be beamformed into distinct distances different from those of the desired signal, while the approximation of Theorem 1 are very tight.

\subsection{Achievable Sum Rates}

Next, we examine the impact of nonlinear distortion on the sum rate as a function of the signal-to-noise ratio (SNR), defined as $\text{SNR}\triangleq \rho/\sigma_n^2$, where $\rho$ represents the received signal power, and $\sigma_n^2$ denotes the noise power at the receiver. Assuming a LoS channel with no path loss, the received power equals the total transmit power at the BS after amplification, which is considered constant. Under this assumption, we derive the achievable sum rates for MRT and ZF precoding using the following expression:
\begin{align}
    R_{\tt sum} = \frac{1}{S}\sum_{k=1}^K\sum_{\nu\in S}\log_2\left( 1 + \gamma_{k}[\nu] \right).
\end{align}
where $\gamma_{k}[\nu]$ is the signal-to-interference-noise-and-distortion ratio (SINDR), defined as:
\begin{align}
    \gamma_{k}[\nu]=\frac{|\veca_k^T \matG \matP_{\nu,k}|^2}{\sum_{i\neq k}|\veca_k^T \matG \matP_{\nu,i}|^2+\veca_k^T\matS_{\tt dd}[\nu] \veca_k^{\ast}+\sigma_n^2},
\end{align}
where $\matP_{\nu,k}$ denotes the $k$-th column of the precoder $\matP_\nu$ (corresponding to the $k$-th user) as defined in \eqref{eq:mrt} and \eqref{eq:zf}.
Specifically, Fig. \ref{fig:sum_rate_16} illustrates the sum rate as a function of the SNR for different levels of EVM (i.e., $0$ and $3\%$). 
When the nonlinear distortion is zero (i.e., EVM $= 0$), the sum rate increases linearly with the SNR for ZF,  as it effectively cancels inter-user interference. However, when introducing an EVM of $3\%$ results in noticeable degradation in the sum rate at high SNR due to the presence of nonlinear distortion, which cannot be mitigated by ZF.
Conversely, MRT precoding does not mitigate inter-user interference, which dominates the distortion. Thus, the performance degradation is negligible in the presence of nonlinear distortion.
Fig. \ref{fig:sum_rate_100} shows the sum rate as a function of SNR with a system equipped with $M=100$ antennas. Notice that the introduction of nonlinear distortion results in a significant reduction in the sum rate, especially at higher SNR values. With $M=100$ antennas, the performance degradation due to distortion is more pronounced than in the $M=16$ case, demonstrating that the higher the number of antennas, the greater the level of distortion at the UEs.
These results clearly demonstrate the impact of EVM on system performance, emphasizing the importance of minimizing EVM to achieve higher sum rates, particularly in scenarios with a larger number of antennas and higher SNR values. The analysis shows that while ZF precoding can handle inter-user interference effectively under ideal conditions, its performance degrades significantly in the presence of nonlinear distortion. On the other hand, MRT precoding shows resilience to nonlinear distortion, but its overall performance is limited by inter-user interference.

\begin{figure}[h]
  \centering
  \includegraphics[width=\columnwidth]{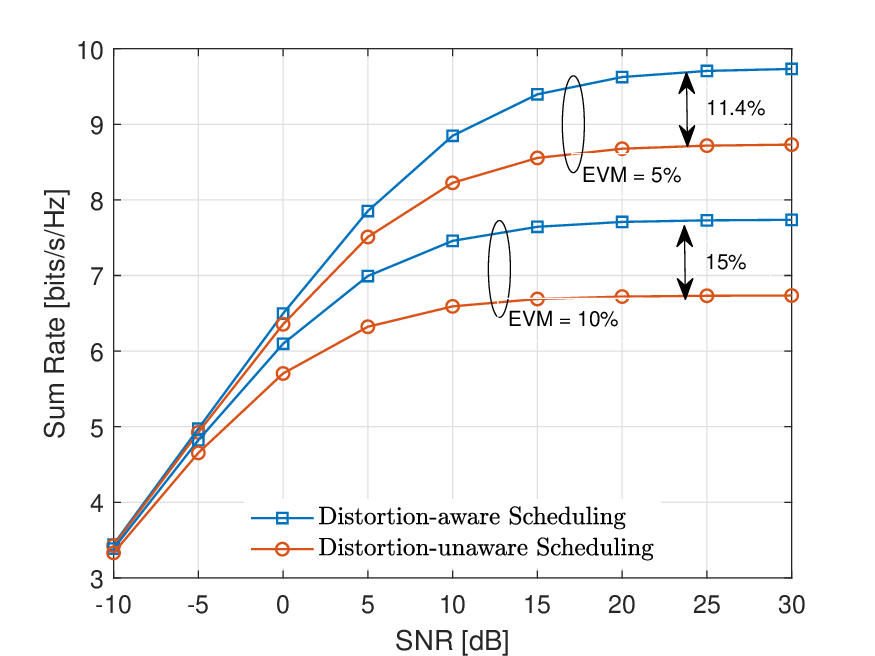}
  \caption{Sum rate as a function of SNR for different levels of EVM. In this example, 4 UEs are co-scheduled employing MRT precoding and sub-band scheduling with $M = 100$ antennas.}
  \label{fig:scheduling}
\end{figure}

\subsection{Distortion-Aware Frequency Scheduling}

It is possible to use Theorem 1 to predict the beamforming directions of the distortion created by nonlinear amplifiers. This insight can be applied to schedule users in the frequency domain to minimize the influence of distortion. To demonstrate this, we consider a scenario where users are grouped into four clusters, with each cluster containing closely located users, while the clusters themselves are physically far apart. In Fig.~\ref{fig:scheduling}, 4 users are co-scheduled in a bandwidth of 128 subcarriers, with each user occupying 30 subcarriers. Sub-band scheduling is employed, meaning that different users are allocated distinct parts of the bandwidth.

In the distortion-unaware scheduling method, users are scheduled randomly. This approach often results in co-scheduling users from the same cluster, exposing them to stronger out-of-band (OOB) emissions due to their physical proximity. These emissions exhibit the highest levels of power in the vicinity of the user (cf. Corollary 1), leading to increased distortion and reduced performance (see Figs. \ref{figure:azimuthSimulation} and \ref{figure:elevationSimulation}).

Conversely, the distortion-aware scheduling method leverages Theorem 1 to predict the beamforming directions of distortion created by nonlinear amplifiers. This enables the minimization of OOB emissions by scheduling users in a way that avoids interference. In this example, by preventing the co-scheduling of users from the same cluster, the method exploits the significant physical separation between clusters to reduce OOB distortion effects that would otherwise be beamformed toward users within the same cluster.

Fig.~\ref{fig:scheduling} presents the sum rate as a function of SNR for the two scheduling methods under different EVM levels. The hardware-aware scheduling method achieves a significant sum rate gain compared to the hardware-unaware method, particularly at high SNR values. Moreover, the gain of the hardware-aware method increases with higher non-linear distortion (higher EVM). For instance, at high SNR, the gain is larger for EVM = 10\% compared to EVM = 5\%, as higher non-linear distortion amplifies the benefits of distortion-aware scheduling. At low SNR values, the performance gap between the two methods is negligible, as non-linear distortion effects are dominated by thermal noise.

\section{Conclusion}

In this paper, we analyzed how ELAAs and active RISs with nonlinear amplifiers introduce distortion in radiative near-field channels. We examined its spatial characteristics across azimuth, elevation, and depth and derived a theorem predicting its beamforming and focusing behavior. While distortion aligns with the desired signal directions, it also appears at distinct near- and far-field locations when co-scheduled users are at different near-field distances.

We further demonstrated that when multiple users are distributed across different elevation angles, distortion spreads into new azimuth and elevation directions, while user positions in azimuth do not affect distortion in the elevation domain. Additionally, nonlinear amplification causes both in-band and out-of-band spectral expansion, increasing interference in adjacent frequencies.

To mitigate these effects, we proposed distortion-aware scheduling based on our theorem, enabling strategic user allocation in space and frequency to reduce interference. This approach significantly improves system performance, particularly in hardware-impaired scenarios, paving the way for more efficient deployment of ELAAs and active RISs.


\appendices
\section{Proof of Theorem 1}\label{appendix:theorem1}

To determine the directions and range of the $p$-th order and the $l$-th memory tap for the non-linear distortion, we express the radiation pattern of the distortion in terms of the array factor, i.e.,
\begin{align} \label{eq:array_factor}
\begin{split}
       {\tt AF}_{2p+1}^{(l)}(\varphi,\theta,r) = \beta_{2p+1}[l] \sum_{m=1}^{M}a_m(\varphi,\theta,r)\\
       \times x_m[n-l]|x_m[n-l]|^{2p}, 
\end{split}
\end{align}
where $a_m(\cdot,\cdot,\cdot)$ is the LoS channel between the $m$-th antenna element and the point $(\varphi,\theta,r)$. 
Furthermore, in order to derive tractable analytical expressions for the beamforming directions and range of distortion, we approximate the channel (a.k.a. Fresnel approximation) as  
\begin{align} 
    a_m(\varphi, \theta, r) &\approx e^{j\frac{2\pi}{\lambda}\tilde{\phi}_m(\varphi, \theta, r)},\label{eq:fresnel_apprx}
\end{align}
which is tight when the propagation distance is larger than the $d_B = 2\Delta$ distance, where $\Delta$ is the maximum linear dimension of the array and the amplitude variations between the point $(\varphi,\theta,r)$ and the aperture of ELAA are negligible\cite{emil_pwr_scale,alva}.
Now, by substituting (\ref{elaa:bb_tx_signal})  into $x_m[n-l]|x_m[n-l]|^{2p}$, we get the expression
\begin{align}
\begin{split}\label{eq:nonlinear-factor}
    x_m[n-l]\left|x_m[n-l]\right|^{2p} = \left( x_m[n-l]\right) ^{p+1}\left( x_m^{\ast}[n-l]\right)^{p}\\
    = \sum_{k_0=1}^{K}\sum_{k_2=1}^{K}\cdots \sum_{k_{2p}=1}^{K}  
    s_{k_0}[n-l]  s_{k_2}[n-l]\ldots s_{k_{2p}}[n-l]\\
    \times e^{-j\frac{2\pi}{\lambda} \sum_{i=0}^p \phi_{m}\left(\varphi_{k_{2i}},\theta_{k_{2i}},r_{k_{2i}} \right)}\\
    \times \sum_{k_{1}=1}^{K}\sum_{k_{3}=1}^{K}\cdots \sum_{k_{2p-1}=1}^{K}  
    s^{\ast}_{k_1}[n-l] s^{\ast}_{k_3}[n-l] \ldots s^{\ast}_{k_{2p-1}}[n-l] \\
    \times e^{j\frac{2\pi}{\lambda} \sum_{i=1}^{p} \phi_{m}\left(\varphi_{k_{2i-1}},\theta_{k_{2i-1}},r_{k_{2i-1}} \right) }\\
    = \sum_{k_{0}=1}^{K}\cdots \sum_{k_{2p}=1}^{K}
     \underbrace{s_{k_{0}}[n-l]s^{\ast}_{k_{1}}[n-l]\cdots  s_{k_{2p}}[n-l]}_{\triangleq 
 s_{{\boldsymbol k}}[n-l]}\\
    \times e^{-j\frac{2\pi}{\lambda}  \sum_{i=0}^{2p} (-1)^{i} \phi_{m}\left(\varphi_{k_{i}},\theta_{k_{i}},r_{k_{i}} \right) }.
 \end{split}
\end{align}
Moreover, for notational convenience, we define the phase of the $p$-th order of nonlinear distortion in \eqref{eq:nonlinear-factor} as
\begin{align}
    \tilde{\phi}_m(\varphi_{\boldsymbol k},\theta_{\boldsymbol k},r_{\boldsymbol k}) \triangleq \sum_{i=0}^{2p} (-1)^{i} \phi_{m}\left(\varphi_{k_{i}},\theta_{k_{i}},r_{k_{i}} \right)
\end{align}
and it can be re-written in the form of expression \eqref{eq:phase_apprx} as follows:
\begin{align}
    \begin{split}\label{eq:nonlinear-phase}
     \tilde{\phi}_m(\varphi_{\boldsymbol k},\theta_{\boldsymbol k},r_{\boldsymbol k})
        = k_z(m)\underbrace{\sum_{i=0}^{2p} (-1)^{i} \sin\theta_{k_i}}_{\triangleq \sin\theta_{\boldsymbol k}}\\
        +k_y(m)\underbrace{\sum_{i=0}^{2p} (-1)^{i} \sin\varphi_{k_i}\cos\theta_{k_i}}_{\triangleq \sin\varphi_{\boldsymbol k}\cos\theta_{\boldsymbol k}}\\
        -\frac{k^2_z(m)+k^2_y(m)}{2} \underbrace{\sum_{i=0}^{2p} (-1)^{i} \frac{1}{r_{k_i}}}_{\triangleq \frac{1}{r_{\boldsymbol k}}}.
    \end{split}
\end{align}
Then by substituting \eqref{eq:nonlinear-factor} into \eqref{eq:array_factor}, the array factor of the $p$-th order and the $l$-th memory tap becomes
\begin{align}
    \begin{split}\label{eq:array_factor_3}
    {\tt AF}_{2p+1}^{(l)}(\varphi,\theta,r)  \approx  \beta_{2p+1}[l] \sum_{k_{0}=1}^{K}\cdots \sum_{k_{2p}=1}^{K}   
    s_{\boldsymbol k}[n-l]\\
    \times \underbrace{  \sum_{m=1}^{M} e^{-j\frac{2\pi}{\lambda} \left(  \tilde{\phi}_m(\varphi,\theta,r) - \tilde{\phi}_m(\varphi_{\boldsymbol k},\theta_{\boldsymbol k},r_{\boldsymbol k})    \right) }}_{\triangleq  
 {\tt AF}_{\boldsymbol k}(\varphi,\theta,r)},
 \end{split}
\end{align}
where the effective azimuth $\varphi_{\boldsymbol k}$, elevation $\theta_{\boldsymbol k}$ and range $r_{\boldsymbol k}$ of the nonlinear distortion are given Theorem \ref{th:nld_char}.
Now, the amplitude of the term ${\tt AF}_{\boldsymbol k}(\varphi,\theta,r)$ can be upper bounded using the generalized triangle inequality as
\begin{align*}
    |{\tt AF}_{\boldsymbol k}(\varphi,\theta,r)| &\le \sum_{m=1}^{M}\left|  e^{-j\frac{2\pi}{\lambda} \left(  \tilde{\phi}_m(\varphi,\theta,r) - \tilde{\phi}_m(\varphi_{\boldsymbol k},\theta_{\boldsymbol k},r_{\boldsymbol k})\right)}\right|\\
    & =  M
\end{align*}
Therefore, the equality  holds if and only if the channel vector equals that of the effective precoder, i.e.,
\begin{align}
    \tilde{\phi}_m(\varphi,\theta,r) =  \tilde{\phi}_m(\varphi_{\boldsymbol k},\theta_{\boldsymbol k},r_{\boldsymbol k})
\end{align}
Therefore, the term $|{\tt AF}_{\boldsymbol k}(\varphi, \theta,r)|$ is maximized at the points where $(\varphi, \theta,r) = (\varphi_{\boldsymbol k}, \theta_{\boldsymbol k},r_{\boldsymbol k})$ for $k_i = 1,2,...,K$ and $i=0,1,...,2p$.

\section{Proof of Corollary 1}
The expression (\ref{eq:array_factor_3}) for the third-order memoryless non-linear distortion can be re-written as
\begin{align}
\begin{split}\label{eq:array_factor_cor}
    {\tt AF}_3(\varphi,\theta,r) = (2K-1){\sum_{p=1}^K c_{ppp} {\tt AF}_{ppp}(\varphi,\theta,r)}\\
         + \sum_{p=1}^{K} \sum\limits_{\substack{q=1 \\ p\neq q}}^K c_{pqp} {\tt AF}_{pqp}(\varphi,\theta,r)\\
         + 2 \sum_{p=1}^{K} \sum\limits_{\substack{q=1 \\ q\neq p}}^K \sum\limits_{\substack{v=1 \\ v\neq p \\ v>q}}^K c_{pqv} {\tt AF}_{pqv}(\varphi,\theta,r)
\end{split}
\end{align}
where
\begin{align}
    c_{pqv} \triangleq \beta_3 s_p[n]s_q^{\ast}[n]s_v[n]
\end{align}
Notice, that each term in (\ref{eq:array_factor_cor}) represents a unique radiation pattern stemming from nonlinear distortion. For example, the first, second and third term yields the focusing points for $\mathcal{P}_1$, $\mathcal{P}_2$ and $\mathcal{P}_3$, respectively.

\ifCLASSOPTIONcaptionsoff
  \newpage
\fi



%

\bibliographystyle{IEEEtran}
\bibliography{bibtex/IEEEabrv,bibtex/confs-jrnls,bibtex/publishers,bibtex/referbib}
\end{document}